\documentclass{pasj01}
\Received{$\langle$reception date$\rangle$}
\Accepted{$\langle$acception date$\rangle$}
\Published{$\langle$publication date$\rangle$}

\begin{document}

\title{Observational Study on the Fine Structure and Dynamics of a Solar Jet. \\
II. Energy Release Process Revealed by Spectral Analysis.}
\author{Takahito \textsc{Sakaue}\altaffilmark{1}, Akiko \textsc{Tei}\altaffilmark{1}, Ayumi \textsc{Asai}\altaffilmark{1}, Satoru \textsc{Ueno}\altaffilmark{1}, Kiyoshi \textsc{Ichimoto}\altaffilmark{1}, and Kazunari \textsc{Shibata}\altaffilmark{1}}%
\altaffiltext{1}{Kwasan and Hida Observatories, Kyoto University, Kyoto, Japan}
\email{sakaue@kwasan.kyoto-u.ac.jp}

\KeyWords{Sun: chromosphere --- Sun: flares  --- Sun: magnetic fields }

\maketitle

\begin{abstract}
We report a solar jet phenomenon associated with the C5.4 class flare on 2014 November 11. The data of jet was provided by Solar Dynamics Observatory (SDO), X-Ray Telescope (XRT) aboard Hinode, Interface Region Imaging Spectrograph (IRIS) and Domeless Solar Telescope (DST) at Hida Observatory, Kyoto University. These plentiful data enabled us to present this series of papers to discuss the entire processes of the observed phenomena including the energy storage, event trigger, and energy release. In this paper, we focus on the energy release process of the observed jet, and mainly describe our spectral analysis on the H$\alpha$ data of DST to investigate the internal structure of the H$\alpha$ jet and its temporal evolution. This analysis reveals that in the physical quantity distributions of the H$\alpha$ jet, such as line-of-sight velocity and optical thickness, there is a significant gradient in the direction crossing the jet. We interpret this internal structure as the consequence of the migration of energy release site, based on the idea of ubiquitous reconnection. Moreover, by measuring the horizontal flow of the fine structures in jet, we succeeded in deriving the three-dimensional velocity field and the line-of-sight acceleration field of the H$\alpha$ jet. The analysis result indicates a part of ejecta in the H$\alpha$ jet experienced the additional acceleration after it had been ejected from the lower atmosphere. This {\it secondary acceleration} was specified to occur in the vicinity of the intersection between the trajectories of the H$\alpha$ jet and X-ray jet observed by Hinode/XRT. We propose that a fundamental cause of this phenomenon is the magnetic reconnection involving the plasmoid in the observed jet.
\end{abstract}

\section{Introduction}
The ubiquitous and diverse natures of solar jet have been confirmed by the numerous observational studies for past a few decades. They are observed by multi-wavelengths, across a wide range of spatial and temporal scales (see \cite{2016SSRv..201....1R} for a review). 

Such various solar jets are the results from the transient energy release by the magnetic reconnection. Since the magnetic reconnection is the ubiquitous process in the magnetized plasma, the jet phenomena are also ubiquitous in the solar atmosphere [the concept of ``ubiquitous reconnection" \citep{1988ApJ...330..474P,2007Sci...318.1591S}]. \citet{1999Ap&SS.264..129S} developed this concept to the unified model accounting for the diverse nature of the explosive phenomena in the solar atmosphere, including the jets and flares (see also \cite{2011LRSP....8....6S}).\par

According to the above idea, the height of the reconnection site in the solar atmosphere determines the properties of the solar jets. In fact, the reconnection in the corona can produce the magnetically driven hot jet, and often lead to the chromospheric evaporation jet (\cite{2001ApJ...550.1051S}; \cite{2014PhDT....S}; \cite{2003ApJ...593L.133M}). When the reconnection occurs in the upper chromosphere or lower corona, the magnetically driven cool jet tends to be associated with the hot jet (\cite{1995Natur.375...42Y}; \cite{1995SoPh..156..245S}; \cite{1999ApJ...513L..75C}; \cite{2008ApJ...683L..83N}). The reconnection in the middle or lower chromosphere generates the slow mode shocks, which can be a trigger of cool jets through the interaction with the transition region (\cite{2011ApJ...731...43N}; \cite{2013PASJ...65...62T}). 

As mentioned above, the origin of the ubiquitous and diverse natures of solar jet have been clearly demonstrated. Nowadays, therefore, the theoretical studies on the solar jets focus on the dynamics of their internal structure, especially its three-dimensionality, including their helical motion (\cite{1984ChA&A...8..294X}; \cite{1987SoPh..108..251K}; \cite{1994A&A...282..240G}; \cite{1996ApJ...464.1016C}), or recurrent nature (\cite{2012ApJ...760...28S}; \cite{2013A&A...555A..19G}; \cite{2017A&A...598A..41C}). These studies are encouraged by the success of \citet{2008ApJ...673L.211M} in the three-dimensional numerical experiments (3D simulations) of solar jet. There are many observed features of jets which had not been reproduced until the 3D simulations; helical jets (\cite{2014ApJ...789L..19F}; \cite{2015ApJ...798L..10L}; \cite{2015A&A...573A.130P}) and recurrent jets (\cite{2010A&A...512L...2A}; \cite{2010ApJ...714.1762P}).

On the other hand, there are still few studies investigating the internal structures of solar jets on the basis of the spectroscopic observation, although it has a great advantage in measuring the three-dimensional structures of jets. That is mainly because such an observation requires reducing either the field of view or time cadence in contrast to the transient and extensive nature of the jet phenomena. In particular, there are little progress in the observational study on where and how much the ejecta is accelerated since the early works such as \citet{1973PASJ...25..447T}, who revealed that the ejecta moved simply along the ballistic trajectory after the impulsive acceleration. Note that \citet{1983A&A...127..337S} and \citet{2003PASJ...55..503M} tried to measure the three-dimensional velocity field of the jets, but their observations were disadvantageous for investigation of the high velocity components in the jets because the velocity ranges in their spectroscopic observations were limited to 70\,km\,s$^{-1}$ at most. Recently, spectral analyses of the solar jets have been recognized as more important to investigate their fine structures and dynamics. \citet{2012ApJ...759...15M} quantitatively confirmed the coexistence of the thermally and magnetically driven jets, based on the spectroscopic observation by Hinode/EIS. \citet{2014ApJ...790L...4Y} found the multiple shock features in the spectra of jets, and suggested the observed jet was driven by the reconnection-generated shock waves. \citet{2015ApJ...801...83C} reported the homologous helical jets observed by spectroheliograph of IRIS and developed the three-dimensional data-driven simulations to comprehend their underlying drivers.
\par

Fortunately, we succeeded in the co-operative observation of a solar jet between the Domeless Solar Telescope (DST) of Hida Observatory, Kyoto University \citep{1985MmKyo..36..385N}, Interface Region Imaging Spectrograph (IRIS; \cite{2014SoPh..289.2733D}), Hinode \citep{2007SoPh..243....3K}, and Solar Dynamics Observatory (SDO; \cite{2012SoPh..275....3P}). In particular, the spectral data in H$\alpha$ was obtained by the spectroheliograph of DST with the high spatial, temporal and spectral resolution, and wide observing wavelength coverage. That enables us to investigate the three-dimensional velocity field of the observed jet, and its temporal evolution. Moreover, thanks to the continuous full-disk observation by SDO including Atmospheric Imaging Assembly (AIA; \cite{2012SoPh..275...17L}) and Helioseismic and Magnetic Imager (HMI; \cite{2012SoPh..275..207S}), it is also possible to discuss the entire processes leading to the observed jet, from the energy build-up to the event trigger and energy release processes. In the previous study (\cite{2017PASJ...69...80S}, heretofore Paper I), we discussed the energy storage phase of the observed jet and related it to the emergence of the satellite spot in the periphery of a $\delta$-type sunspot. Thus, this paper pays particular attention to the energy release process of the jet.

The article is organized as follows. We explain the overview of our observation, especially the spectroscopic observation by DST in the following section. In section \ref{sec:analysis}, we introduce the analysis methods of the observed spectra (in subsection \ref{sec:spectral analysis by the cloud model}) and the horizontal velocity field of the jet (in subsection \ref{sec:Analysis on the Horizontal Velocity by NAVE}), and provide the results of them. The additional analyses on the observed fine structure of jet are presented in subsection \ref{sec:Investigation of the Secondary Acceleration Region}. Finally, based on those results, we discuss the internal structure and three-dimensionality of the observed jet in section \ref{sec:discussion}.

\begin{figure*}[t]
\begin{center}
\FigureFile(85mm,45mm){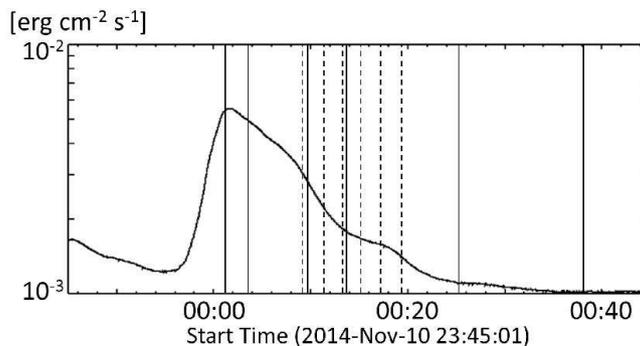}
\caption{The soft X-ray light curve observed by GOES. The peak is C5.4, at 00:01UT. The solid lines indicate the times of panels in figures \ref{fig:ha+aia304time_sequence}, \ref{fig:aia193+094time_sequence}, and the dash lines correspond to those of figure \ref{fig:xrt_time_sequence}.}
\label{fig:goes_for_time_sequence}
\end{center}
\end{figure*}
\begin{figure*}[p]
\begin{center}
\FigureFile(150mm,230mm){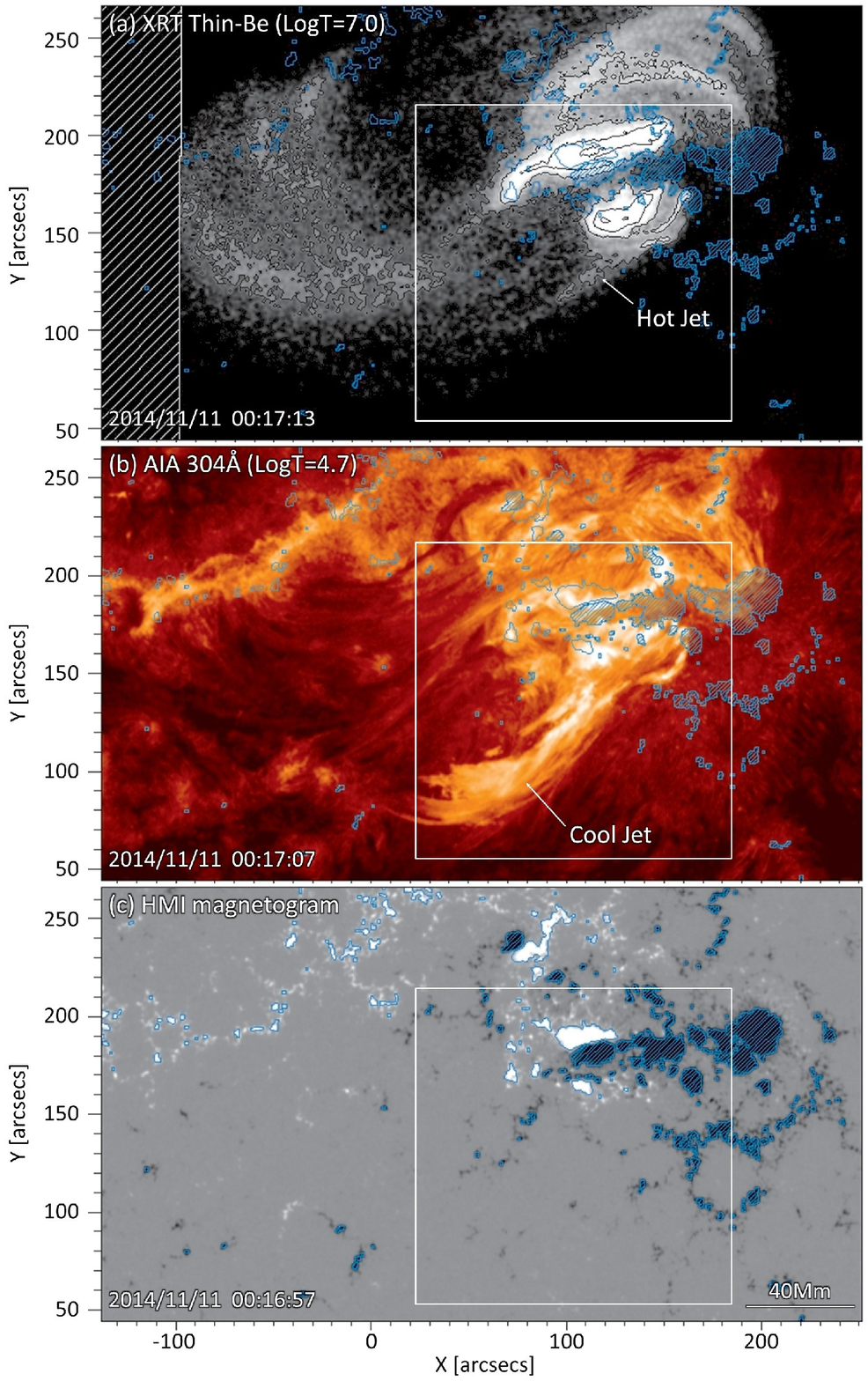}
\caption{The simultaneous snapshots of the jet in (a) Hinode/XRT, (b) SDO/AIA 304\AA\ images and (c) SDO/HMI line-of-sight magnetogram. The observed jet consisted of the cool ($\sim10^{4-5}$K) and hot ($\sim10^{6-7}$K) components. The black contour in XRT image shows its brightness distribution, and blue one in each image represents the regions where the absolute magnetic field strength is larger than 200G; also, the negative polarities are shaded. The white squares in this figure represent the fields of view of the images in figures \ref{fig:ha+aia304time_sequence}--\ref{fig:xrt_time_sequence}. (Color Online and animation of AIA 304\AA\ images is available at the URL\footnotemark[1].)}
\label{fig:jet_snapshots}
\end{center}
\end{figure*}
\footnotetext[1]{$\langle$http://www.kwasan.kyoto-u.ac.jp/~sakaue/20141111/AIA304\_jet.m1v$\rangle$}

\begin{figure*}[p]
\begin{center}
\FigureFile(165mm,110mm){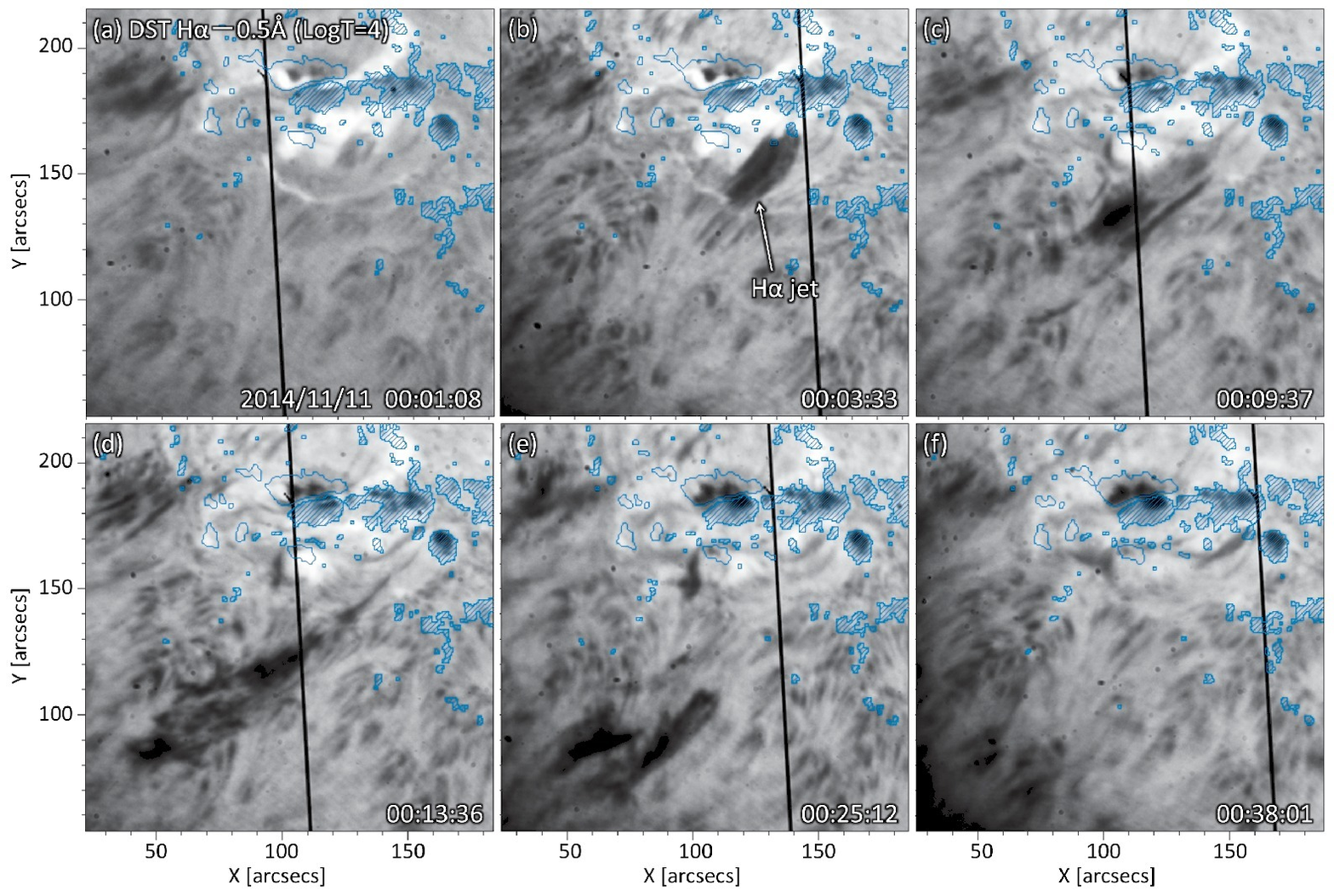}
\FigureFile(165mm,110mm){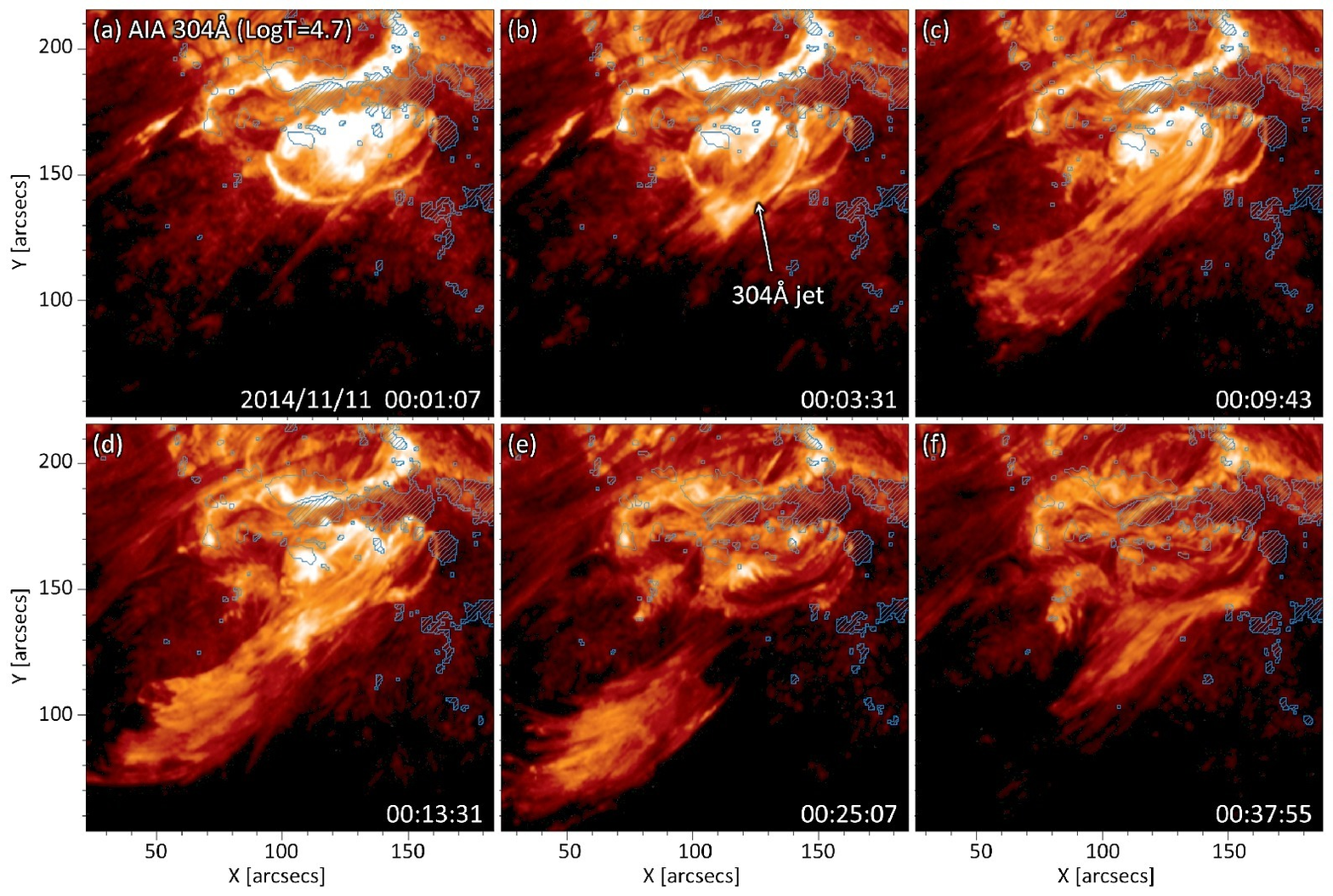}
\caption{Time sequences of DST H$\alpha$-0.5\AA\ slit-jaw images (top) and SDO/AIA 304\AA\ images (bottom). The field of view is indicated with the white squares in figure \ref{fig:jet_snapshots}. The contours in these images represent the regions where the absolute magnetic field strength is larger than 200G; also, the negative polarities are shaded. The observed jet emanated after the soft X-ray flux observed by GOES reached the peak, C5.4, at 00:01UT (a). The ejection of jet lasted for ten minutes ((b)--(d)), and gradually fell down ((e)--(f)). (Color online)}
\label{fig:ha+aia304time_sequence}
\end{center}
\end{figure*}
\begin{figure*}[tbp]
\begin{center}
\FigureFile(165mm,110mm){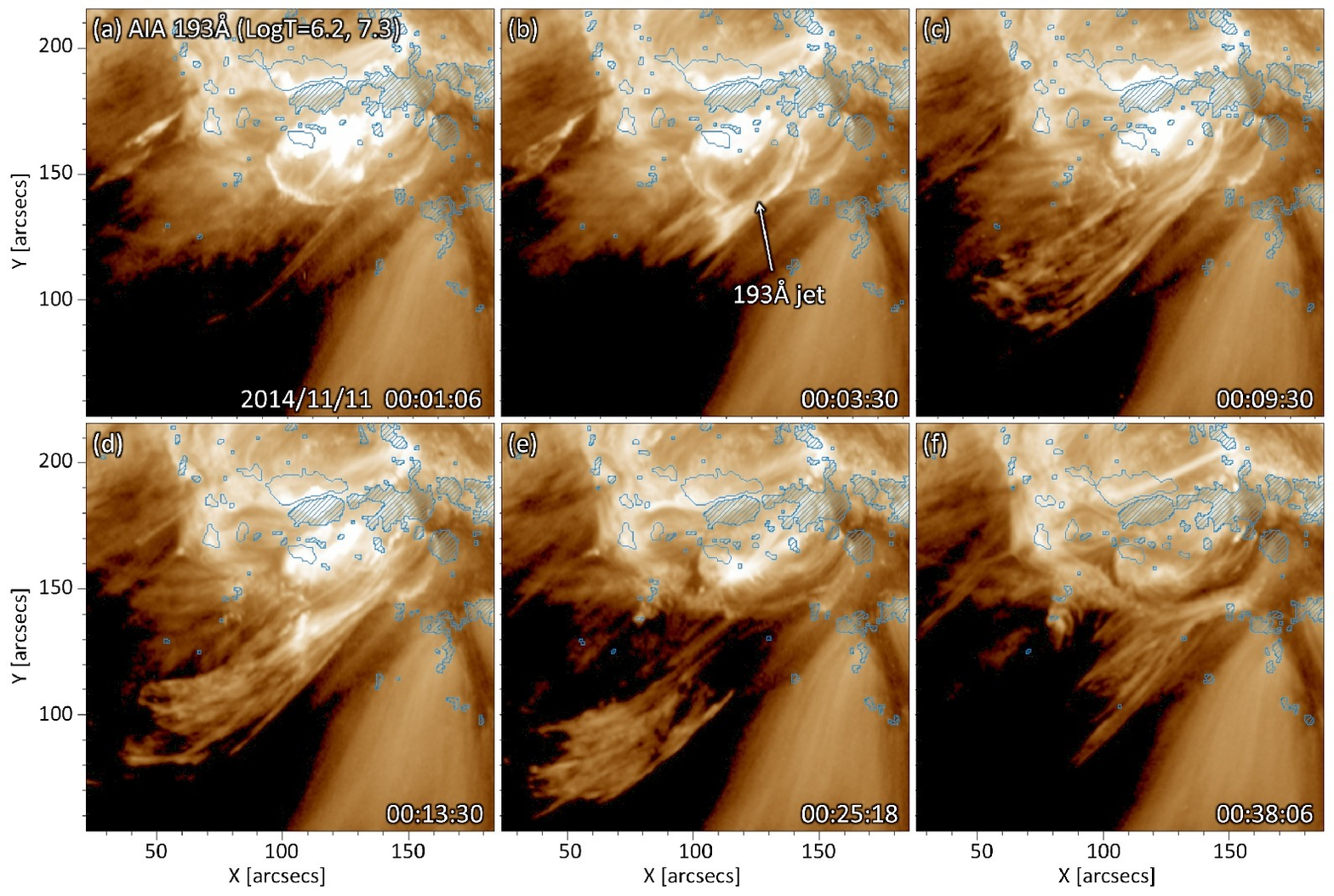}
\caption{Time sequence of AIA 193\AA. The field of view is the same as that of figure \ref{fig:ha+aia304time_sequence}. Compared to the previous time sequences, the hotter plasma ($T>10^6$K) in the jet is displayed here. (Color online)}
\label{fig:aia193+094time_sequence}
\FigureFile(165mm,110mm){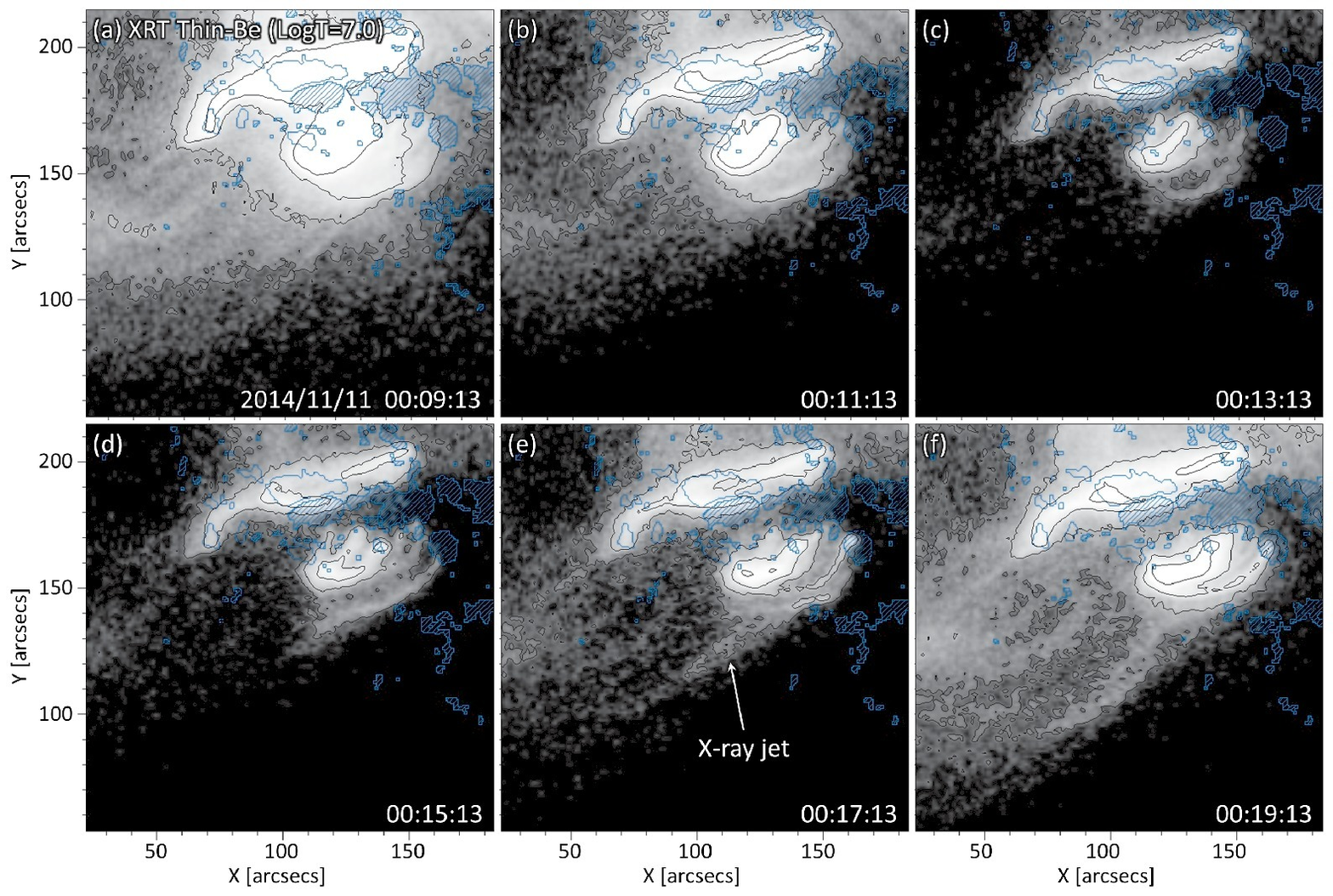}
\caption{Time sequence of XRT images. The field of view is the same as that of figures \ref{fig:ha+aia304time_sequence}, \ref{fig:aia193+094time_sequence}. The black contours show the X-ray brightness distribution. The X-ray jet emanated about 14 minutes after the H$\alpha$ jet ejection set in (figure \ref{fig:ha+aia304time_sequence}). The GOES soft X-ray flux at each time of those images is indicated in figure \ref{fig:goes_for_time_sequence} with dash lines. (Color online)}
\label{fig:xrt_time_sequence}
\end{center}
\end{figure*}

\begin{figure*}[tbp]
\begin{center}
\FigureFile(170mm,944mm){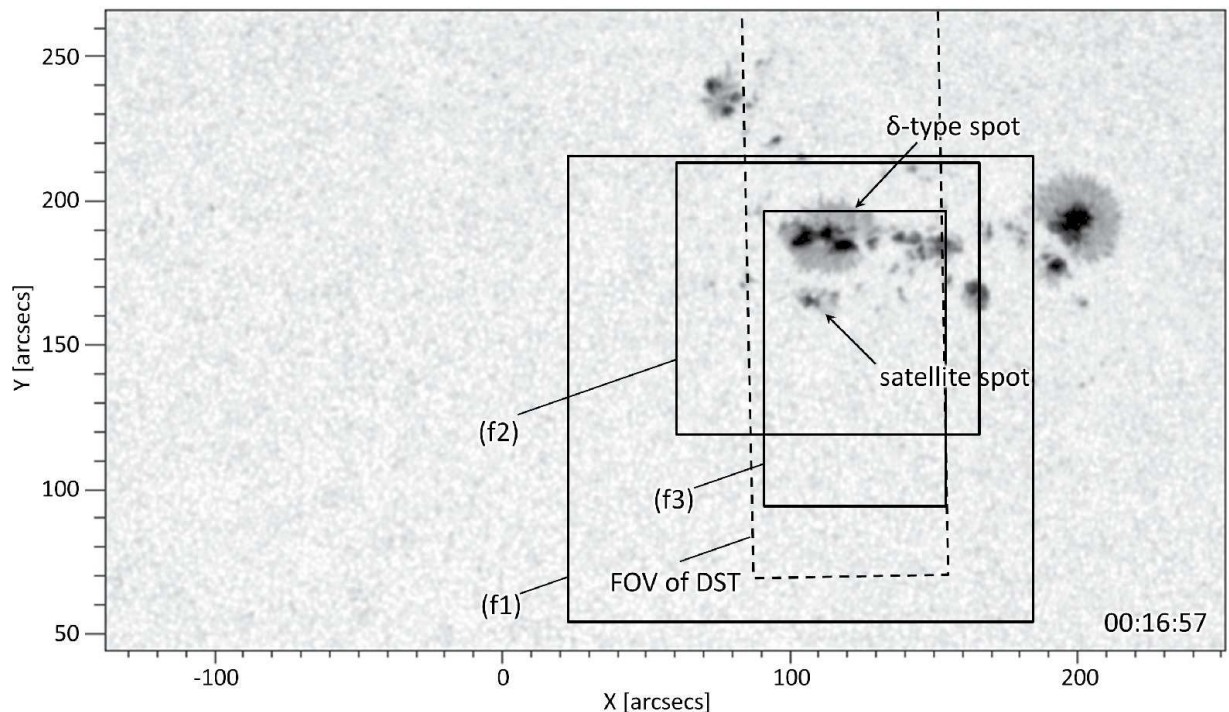}
\caption{HMI continuum image at 00:16:57. The observed jet was triggered as the result of the emergence of the satellite spot near the $\delta$-type spot (see Paper I for the details). The dash-line frame corresponds to the field of view of the spectroheliogram taken by the DST, while the solid-line frames, labeled (f1), (f2) and (f3) represent those of figures \ref{fig:ha+aia304time_sequence}--\ref{fig:xrt_time_sequence}, figure \ref{fig:jet_timeslice_iris}, and figures \ref{fig:vlos_ev}, \ref{fig:vlos_grad}, \ref{fig:hv_losv}, \ref{fig:sar}, \ref{fig:2com_map}, respectively.}
\label{fig:hmi_con_fov}
\end{center}
\end{figure*}
\begin{figure*}[tbp]
\begin{center}
\FigureFile(170mm,112mm){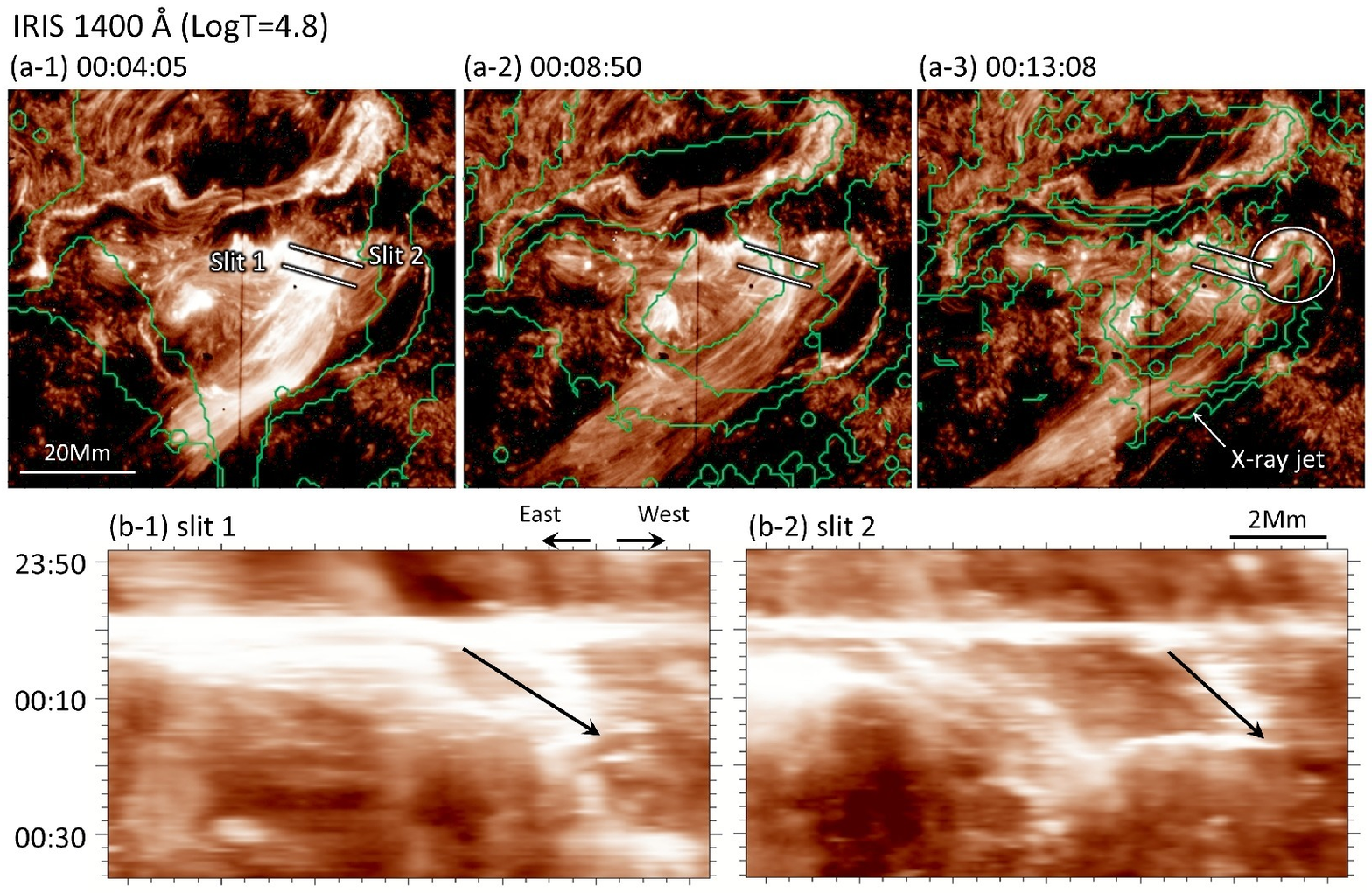}
\caption{(a): Time sequence of IRIS 1400\AA\ images, which shows the temporal evolution of the cool jet ($\sim 10^{4.8}$K). Its field of view is indicated with the solid-line frame labeled (f2) in figure \ref{fig:hmi_con_fov}. The green contours represent the X-ray brightening features observed by XRT, especially X-ray jet seen in figure \ref{fig:xrt_time_sequence} (panel (a-3)). (b): Timeslice diagrams of IRIS images, the slits for which are drawn in panels (a) as slit1, 2, respectively. The footpoint of jet in IRIS images is seen as the bright threads in these diagrams, and it gradually moved westward, as indicated with black arrows. On the other hand, the footpoint of X-ray jet in XRT images lay to the west of that of cool jet in IRIS images, which is enclosed by the white circle in panel (a-3).  (Color online)}
\label{fig:jet_timeslice_iris}
\end{center}
\end{figure*}

\section{Observation}
\label{sec:observation}


We analyzed the jet phenomenon which occurred in the active region NOAA 12205 near the disk center (N16W05) on 2014 November 11. This jet was associated with the C5.4 class flare. The temporal variation of the soft X-ray flux of this flare observed by Geostationary Operational Environmental Satellite (GOES) is presented in figure \ref{fig:goes_for_time_sequence}. Owning to the coordinated observation between Hinode, IRIS, and DST of Hida Observatory, as well as to the continuous full disk observation by SDO, this event was successfully captured in each field of view.\par

Figure \ref{fig:jet_snapshots} shows the simultaneous snapshots of the observed jet in (a) the Hinode X-Ray Telescope (XRT; \cite{2007SoPh..243...63G}), (b) 304\AA\ channel of AIA images, and (c) HMI line-of-sight magnetogram. The multi-wavelength observation of AIA and XRT indicates that the observed jet consisted of the cool ($\sim10^{4.7}$K) and hot ($\sim10^{7.0}$K) components.\par

The temporal evolution of this jet in H$\alpha$ ($\sim 10^4$K) by DST, 304\AA\ ($\sim 10^{4.7}$K), 193\AA\ ($\sim 10^{6.2}, 10^{7.3}$K) by AIA, and X-ray ($\sim 10^{7.0}$K) by XRT are shown in figures \ref{fig:ha+aia304time_sequence}--\ref{fig:xrt_time_sequence}. The soft X-ray flux at each time of figures \ref{fig:ha+aia304time_sequence}, \ref{fig:aia193+094time_sequence} is indicated with the solid line in figure \ref{fig:goes_for_time_sequence}, and that of figure \ref{fig:xrt_time_sequence} is indicated with dash line. The fields of view of figures \ref{fig:ha+aia304time_sequence}--\ref{fig:xrt_time_sequence} are the same as each other, and represented by the solid-line squares in figures \ref{fig:jet_snapshots} and \ref{fig:hmi_con_fov}. Figure \ref{fig:hmi_con_fov} is the HMI continuum image corresponding to figure \ref{fig:jet_snapshots}. In paper I, we discuss the emergence of the satellite spot near the $\delta$-type sunspot, seen in that figure, as the energy storage process of the observed jet.
\par

The jet was launched at the same time as the GOES soft X-ray flux reached the peak (C5.4) at 00:01UT (panel (a) of figures \ref{fig:ha+aia304time_sequence}, \ref{fig:aia193+094time_sequence}), and ascended along the closed magnetic loop for the east end of the active region (panels (b)--(d) of figures \ref{fig:ha+aia304time_sequence}, \ref{fig:aia193+094time_sequence}). About 14 minutes after the ejection of cool jet seen in H$\alpha$ and 304\AA, the X-ray jet emanated as seen in figure \ref{fig:xrt_time_sequence}. Figure \ref{fig:jet_timeslice_iris} is presented to investigate the spatial and temporal relationship between the preceding cool jet and the subsequent hot jet. The panels (a-1,2,3) in that figure are the time sequence of IRIS 1400\AA\ images, and show the temporal evolution of the cool jet ($\sim 10^{4.8}$K) with high spatial resolution of 0''.34. Its field of view is indicated with the solid-line frame labeled (f2) in figure \ref{fig:hmi_con_fov}. The green contours in these panels represent the X-ray brightening features observed by XRT. In particular, the X-ray jet appears in panel (a-3), which is also seen in panel (d) of figure \ref{fig:xrt_time_sequence}. The panels (b-1,2) are the timeslice diagrams of IRIS images, the slits for which are drawn in panels (a-1,2,3) as slit1, 2, respectively. The footpoint of jet in IRIS images is seen as the bright threads in the timeslice diagrams, and it gradually moved westward, as indicated with black arrows. The footpoint of X-ray jet, on the other hand, was enclosed by the white circle in panel (a-3), which lay to the west of the mentioned moving footpoint of the cool jet. The panel (a-3) also shows the trajectories of cool and hot jets appear to overlay each other. The ejecta finally fell down about 20 minutes after the onset of the ejection (panels (e)--(f) of figures \ref{fig:ha+aia304time_sequence}, \ref{fig:aia193+094time_sequence}). \par

By the raster scan performed by DST, we obtained the H$\alpha$ spectra of the jet in temporal steps of 10 seconds. The DST's slit swept repeatedly the field of view from west to east with spatial steps of 0".64 and a spatial resolution of 0''.71 along the slit. The spectral sampling of the obtained data is 0.022\AA, corresponding to 2.0\,km\,s$^{-1}$ in the velocity resolution, and its wavelength coverage ranges from -8\AA\ to 8\AA\ around 6562.8\AA\ (H$\alpha$).\par

\begin{figure*}[tbp]
\begin{center}
\FigureFile(150mm,83mm){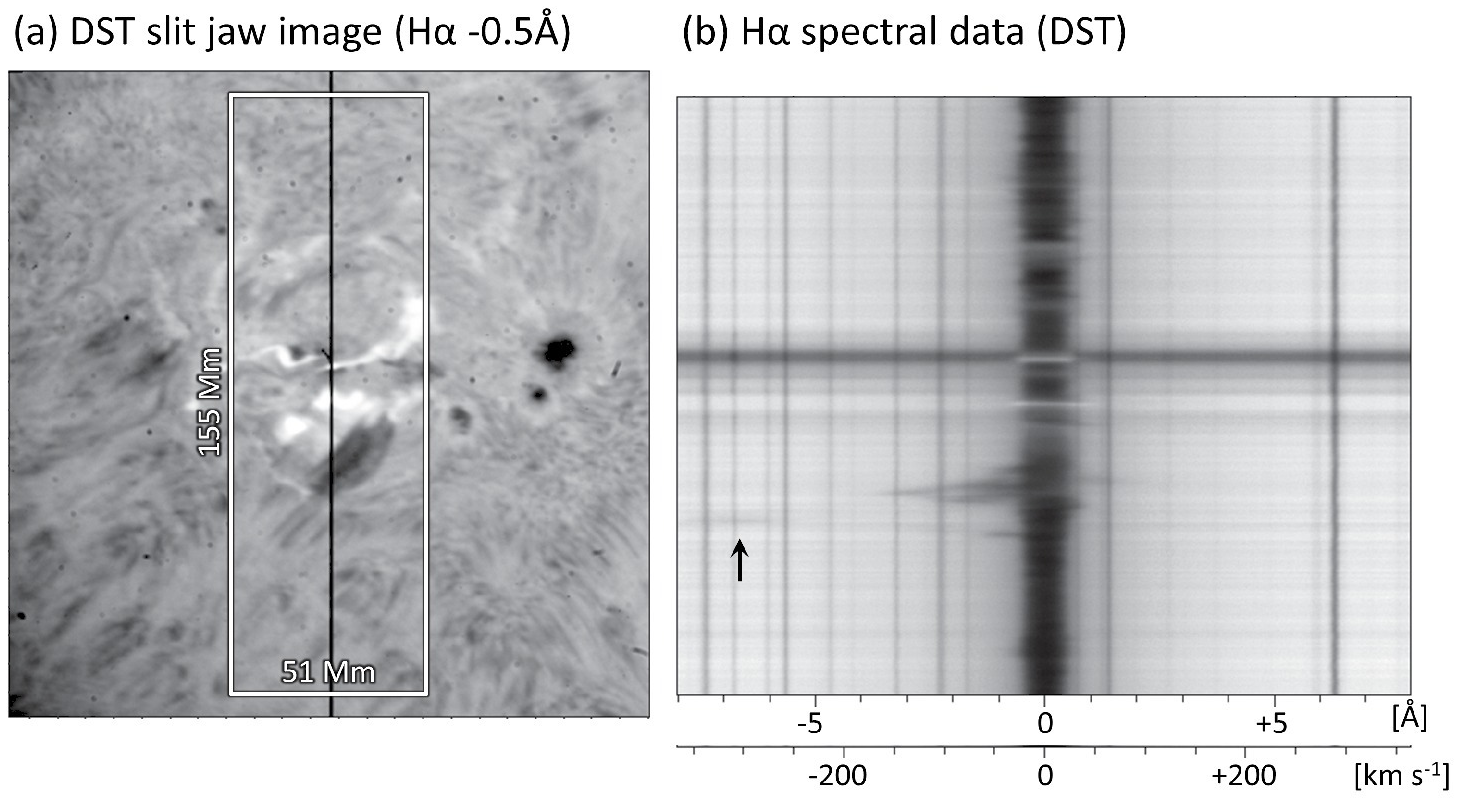}
\caption{(a): The slit jaw image of DST in which the observed jet is imaged. The white rectangle represents the scanning range of our spectroscopic observation (b): The H$\alpha$ spectral image taken at the same time as the panel (a). There is a faint shadow, as pointed at by the arrow, which implies the existence of ejecta with the large upward velocity over 300\,km\,s$^{-1}$.}
\label{fig:dst_his_hs}
\end{center}
\end{figure*}

The panel (a) of figure \ref{fig:dst_his_hs} shows a slit jaw image of DST in which the H$\alpha$ jet is imaged, and the panel (b) shows an H$\alpha$ spectral image taken at the same time. The scanning range of the spectroscopic observation is represented by a white rectangle in the panel (a).\par
It is notable that there is a faint shadow in the spectral image (panel (b)), as pointed at by the arrow, which implies the existence of ejecta with high velocity over 300\,km\,s$^{-1}$. Our analysis started for measuring the line-of-sight velocity accurately from such signals in spectra.

\section{Analysis}
\label{sec:analysis}
\subsection{Spectral Analysis by Using the Cloud Model}
\label{sec:spectral analysis by the cloud model}
\subsubsection{Method}
We applied the cloud model \citep{1964PhDT........83B} to the observed H$\alpha$ spectra of the jet. In this model, the observed spectral profile $I_\lambda$ is interpreted as that composed of the radiation from the background $I_{0\lambda}$, and the emission and absorption of the ejecta itself, so that the contrast of $(I_\lambda-I_{0\lambda})$ to $I_{0\lambda}$ can be written as follows.

\begin{equation}
{I_\lambda-I_{0\lambda}\over I_{0\lambda}}=\left({S\over I_{0\lambda}}-1\right)(1-e^{-\tau_\lambda}) \label{eq:cloud_model}
\end{equation}
\begin{equation}
\tau_\lambda=\tau_0\exp\left[-{1\over2}\left({\lambda/\lambda_0-(1+v_{\mbox{\scriptsize los}}/c)\over W/c}\right)^2\right] \label{eq:tau}
\end{equation}
where $S$, $\tau_0$, $v_{\mbox{\scriptsize los}}$, and $W$ are the source function, optical thickness, line-of-sight velocity and velocity dispersion of the ejecta, respectively, and $c, \lambda_0$ are the speed of light and the wavelength of H$\alpha$, 6562.8\AA. Those four physical parameters of the ejecta ($S$, $\tau_0$, $v_{\mbox{\scriptsize los}}$, and $W$) can be determined so that the observed contrast function (the left side of Eq.\,(\ref{eq:cloud_model})) is reproduced.\par

This model is based on the following three presumptions. The source function of the ejecta is constant against the wavelength and uniform along the line-of-sight direction in the cloud. The absorption coefficient profile of the ejecta is a Gaussian, with a central position displaced by Doppler shift from the line center at rest (see Eq.\,(\ref{eq:tau})). It should be noted that the above contrast function was practically modified for the reason mentioned in the Appendix, where we also explain that there are the cases in which the above equation must be extended to account for the line-of-sight structures of the observed jet more accurately.

\begin{figure*}[tb]
\begin{center}
\FigureFile(170mm,135mm){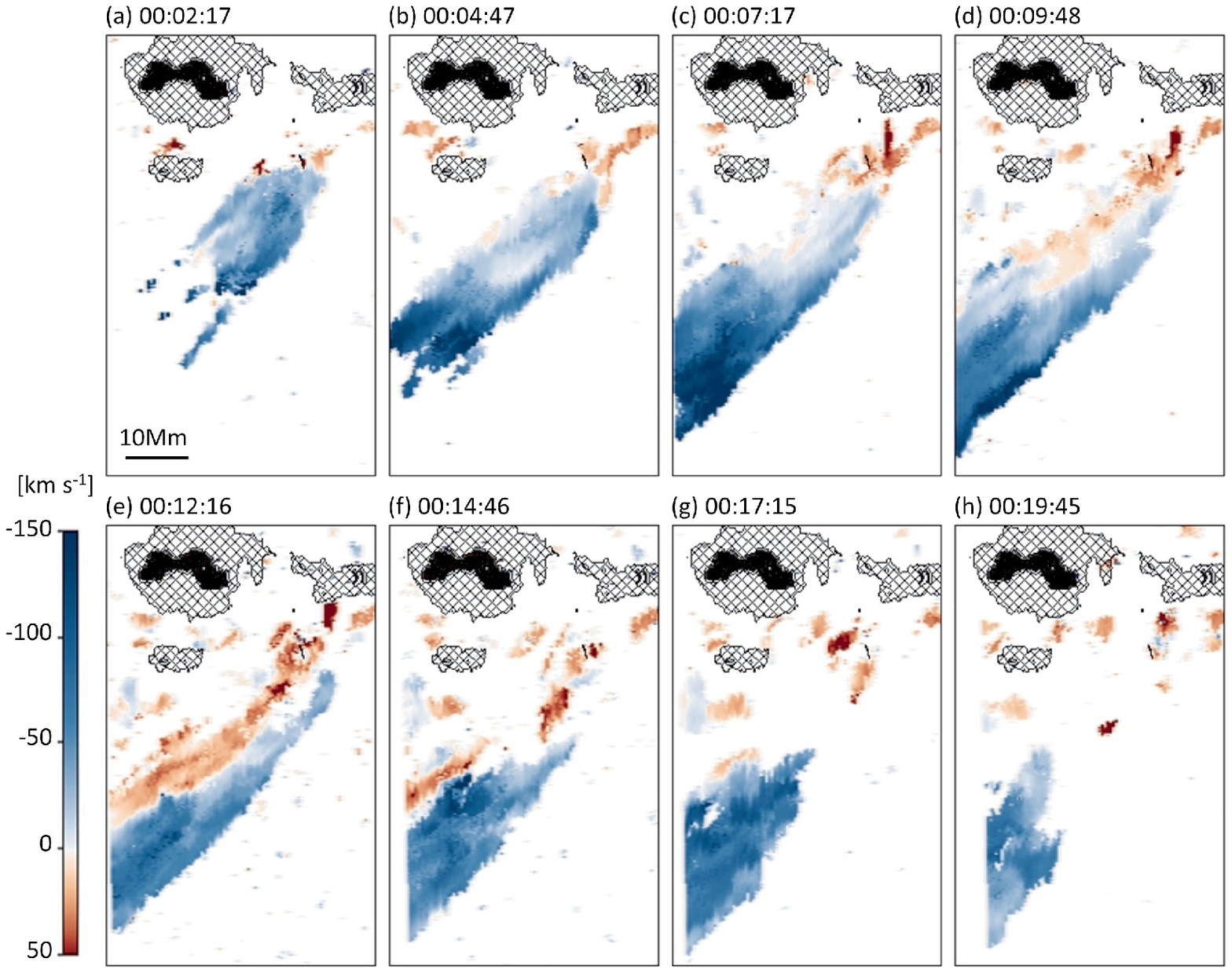}
\caption{The spatial distribution of line-of-sight velocity, and its time sequence in an interval of 150 seconds. The negative direction is upward. The growth of the separation of line-of-sight velocity field is recognized, which means the downward flow developed alongside the upward flow. The shaded and filled area in this figure represent the penumbra and umbra of sunspots, which are defined as the regions whose intensity is smaller than 90\% and 75\% of the continuum level, respectively.}
\label{fig:vlos_ev}
\end{center}
\end{figure*}

\begin{figure*}[tb]
\begin{center}
\FigureFile(138mm,138mm){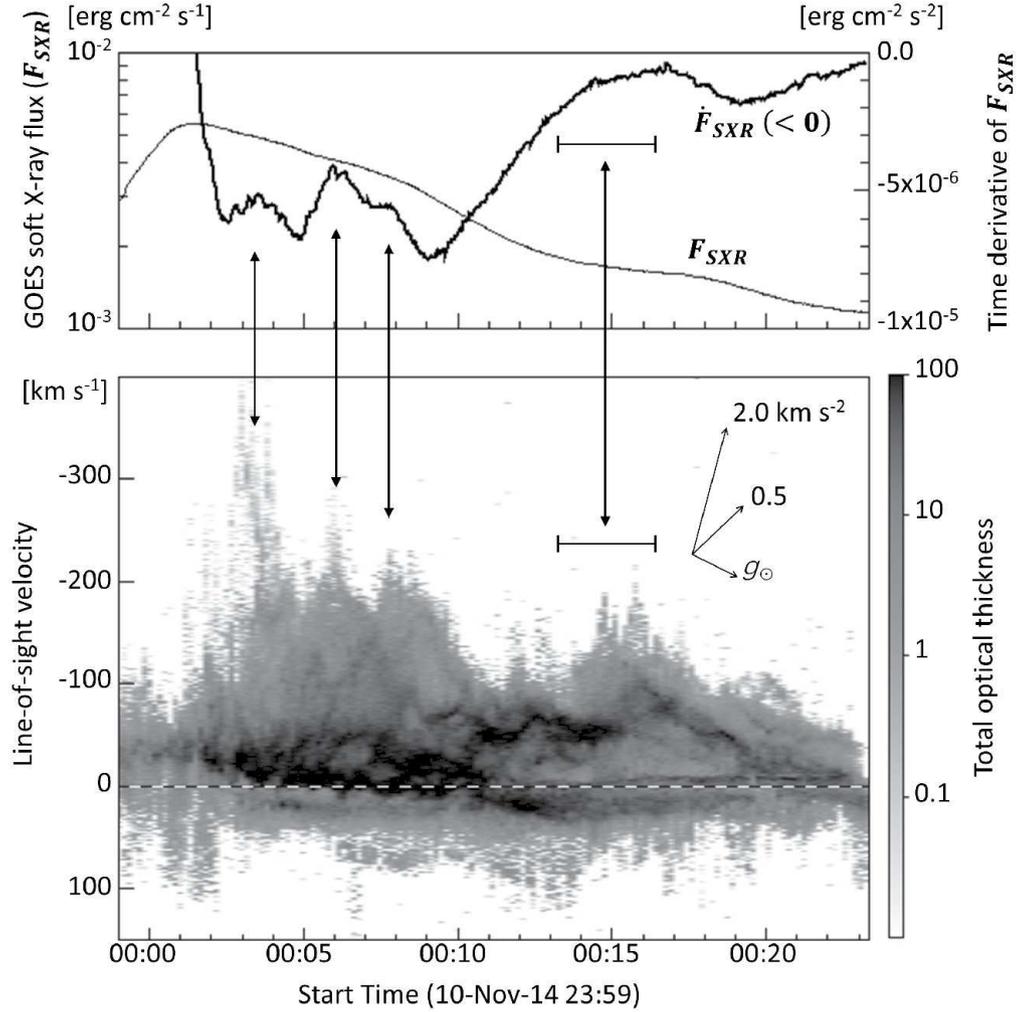}
\caption{{\it Top}: The light curve of soft X-ray flux observed by GOES (thin line) and its time derivative (thick line) during the same period as the bottom of the figure. {\it Bottom}: The temporal variation of the total optical thickness distribution with respect to the line-of-sight velocity. This figure allows us to interpret the temporal evolution of the H$\alpha$ jet as composed of two phases; the primary eruption phase during 00:00 to 00:05, and the secondary acceleration phase indicated by the dimension line from 00:13 to 00:16. It is notable that there is a good correlation between the time variation of line-of-sight velocity with that of time derivative of soft X-ray flux, as indicated by the arrows.}
\label{fig:vlos_time}
\end{center}
\end{figure*}

\begin{figure*}[tb]
\begin{center}
\FigureFile(170mm,83mm){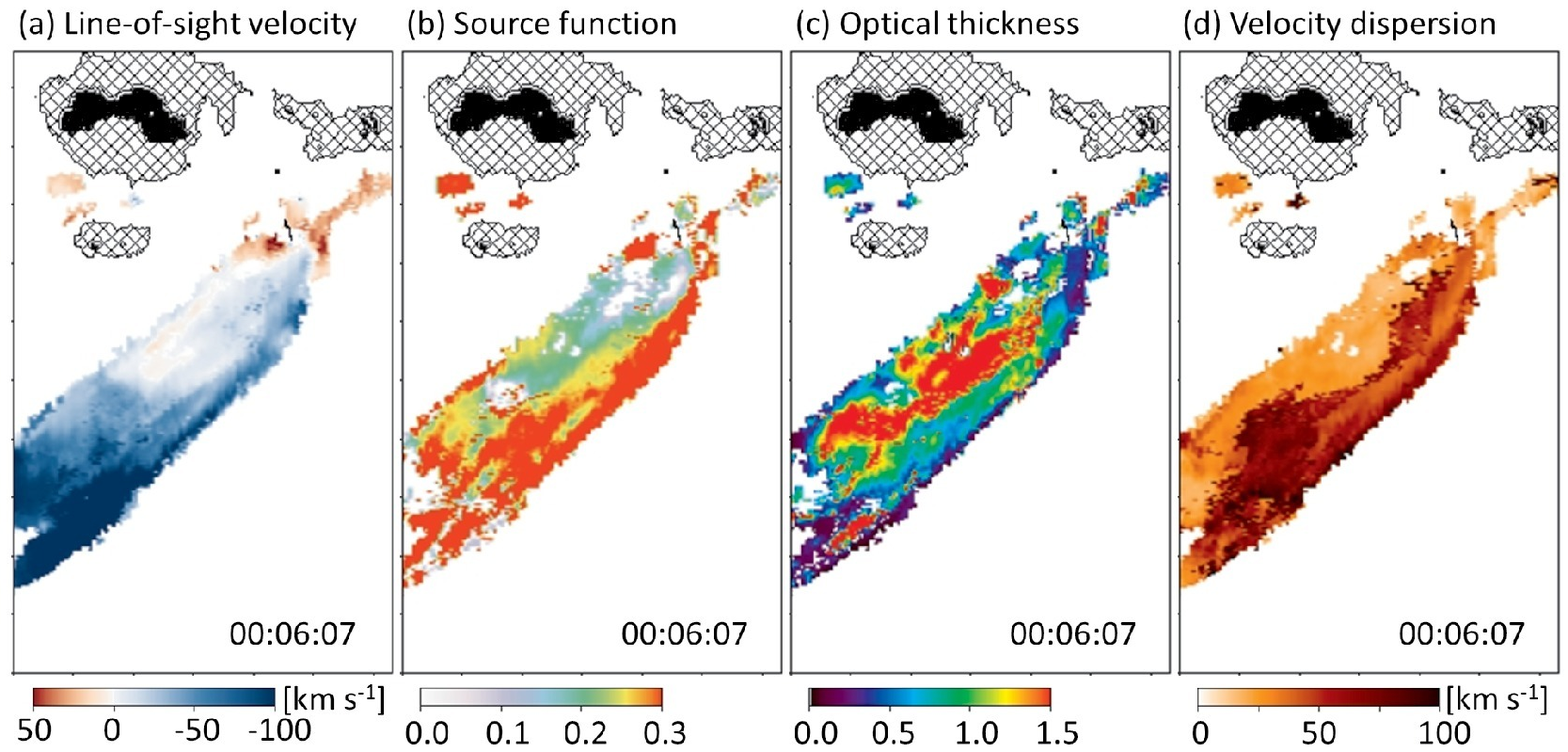}
\caption{The spatial distribution of  the (a) line-of-sight velocity, (b) source function, (c) optical thickness, (d) velocity dispersion of the H$\alpha$ jet at 00:06:07. The source function is normalized with the continuum level. This figure shows that there is a significant gradient in the direction crossing the jet in each physical quantity distribution. In fact, the upward line-of-sight velocity, source function, and velocity dispersion in the right side, as viewed toward the travel direction of the jet, is smaller than in the left side, while the optical thickness in the right side is larger than in the left side. (Color online and animation is available at the URL\footnotemark[2].)}
\label{fig:vlos_grad}
\end{center}
\end{figure*}

\subsubsection{Result}
Figure \ref{fig:vlos_ev} and figure \ref{fig:vlos_time} are the results of our spectral analysis, which show the temporal evolution of the line-of-sight velocity field of the H$\alpha$ jet. Figure \ref{fig:vlos_ev} is the time sequence of the line-of-sight velocity field of the H$\alpha$ jet in an interval of 150 seconds. The negative direction is upward. Figure \ref{fig:vlos_time} shows the temporal variation of the total optical thickness distribution with respect to the line-of-sight velocity (bottom panel), along with the light curve of soft X-ray flux observed by GOES and its time derivative (top panel). The bottom panel can be roughly interpreted as the time variation of the momentum distribution of H$\alpha$ jet. That is correlated with the time derivative of soft X-ray flux observed by GOES, as indicated with the arrows in this figure.\par

Figure \ref{fig:vlos_time} also indicates that the temporal evolution of the H$\alpha$ jet consisted of two phases; the primary eruption phase during 00:00 to 00:05, and the secondary acceleration phase during 00:13 to 00:16. In the primary eruption phase, the line-of-sight velocity of the most accelerated ejecta reached 300\,km\,s$^{-1}$ in 3 minutes, with the acceleration of 1.7\,km\,s$^{-2}$ at least. The largest possible acceleration reaches a few$\times$10\,km\,s$^{-2}$ because the ejecta with 300\,km\,s$^{-1}$ appears abruptly in figure \ref{fig:vlos_time}. Since the maximum line-of-sight velocity is much larger than the typical sound speed in H$\alpha$ jet (10\,km\,s$^{-1}$), the observed H$\alpha$ jet in this phase was probably driven by the Lorentz force, which is able to accelerate the plasma to the Alfv\'en speed; $V_A=B_{\mbox{\scriptsize jet}}/\sqrt{4\pi\rho_{\mbox{\scriptsize jet}}}\sim 280 \mbox{ km s}^{-1}$ ($B_{\mbox{\scriptsize jet}}\sim$10 G, $\rho_{\mbox{\scriptsize jet}}\sim10^{-14}$ g cm$^{-3}$)

On the other hand, it is not self-evident why the secondary acceleration phase appears in figure \ref{fig:vlos_time}. There are two possible causes. First, the eruption mechanism of the jet was activated again around 00:15, and the ejecta with higher velocity was launched from the jet's footpoint. Second, the ejecta already accelerated by 00:15 gained an additional momentum in some way. This phenomenon will be investigated later in detail.\par

In figure \ref{fig:vlos_ev}, the notable feature is that the red shift (downward flow) develops in the right side as viewed toward the travel direction of the jet, while the blue shift (upward motion) continues in the left side. The similar internal structure is shared by the other physical quantities obtained by the spectral analysis. That is to say, in the spatial distribution of each physical quantity, there is a significant gradient in the direction crossing the jet. That appears in figure \ref{fig:vlos_grad}, which shows the spatial distribution of physical quantities at 00:06:07; (a) line-of-sight velocity, (b) source function, (c) optical thickness, (d) velocity dispersion of the H$\alpha$ jet. Note that the source function is normalized with the continuum level. As these panels show, the upward line-of-sight velocity, source function, and velocity dispersion in the right side, as viewed toward the travel direction of the jet, is smaller than in the left side, while the optical thickness in the right side is larger than in the left side.

\footnotetext[2]{$\langle$http://www.kwasan.kyoto-u.ac.jp/~sakaue/20141111/DST\_Ha\_cmp\_m1.5.m1v$\rangle$}

\begin{figure*}[tb]
\begin{center}
\FigureFile(170mm,67mm){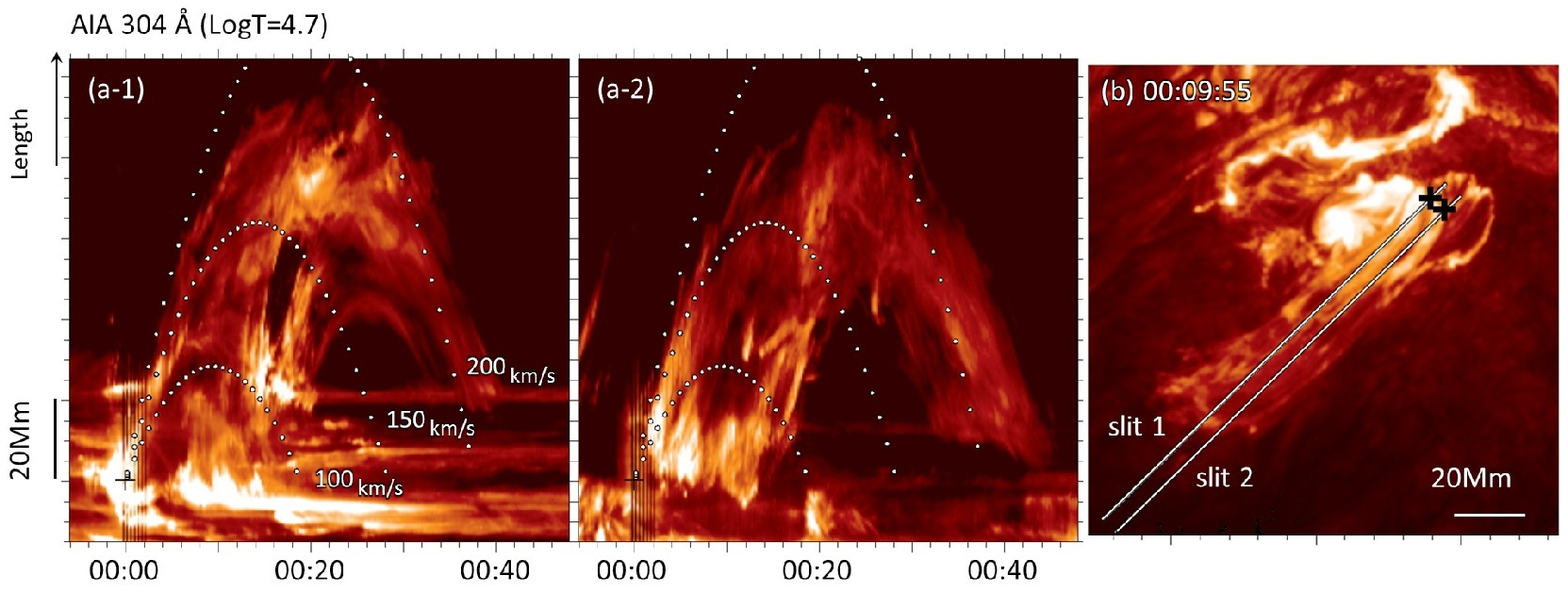}
\caption{Timeslice diagrams of AIA 304\AA\ images along the slits parallel to the jet's travel direction. The two slits, 1 and 2, are drawn in the panel (b), which is the snapshot of AIA 304\AA\ image, and each timeslice diagram is presented in panel (a-1) and (a-2), respectively. The white dot lines in these diagrams represent the ballistic trajectories with several initial velocities (100, 150, 200\,km\,s$^{-1}$) and the same deceleration of 0.176\,km\,s$^{-2}$, corresponding to the case in which the solar surface gravity $g_\odot=0.27$\,km\,s$^{-2}$ and the jet's trajectory inclined at 50 degrees. (Color online)}
\label{fig:jet_timeslice_aia304}
\end{center}
\end{figure*}

\subsection{Optical Flow Analysis for the Horizontal Velocity Field}
\label{sec:Analysis on the Horizontal Velocity by NAVE}
\subsubsection{Method}
In order to measure the horizontal flow of jet and to obtain its three-dimensional velocity field, we applied the optical flow method, called nonlinear affine velocity estimator (NAVE; \cite{2008ApJ...689..593C}), to the sequence EUV images from AIA 304\AA. The reasons why we analyzed the AIA 304\AA\ images rather than H$\alpha$ data is that it is easier to track the dynamic flow of the fine structures in the AIA 304\AA\ images, partly due to their effectively higher spatial resolution with no seeing effect. 
Here, we assume the horizontal flow field measured from the AIA 304\AA\ images is alternative to that of the H$\alpha$ images since their characteristic temperatures are close to each other; a few tens of thousands K. This assumption is also justified by the close similarity of the jet structures seen in both wavelengths. \par

Figure \ref{fig:jet_timeslice_aia304} shows the timeslice diagrams of AIA 304\AA\ images along the slits parallel to the jet's travel direction [panel (a-1, 2)]. The two slits 1 and 2 for them are drawn in the panel (b), which is the snapshot of AIA 304\AA\ image. The white dot lines in these diagrams represent the ballistic trajectories with several initial velocities (100, 150, 200\,km\,s$^{-1}$) and the same deceleration of 0.176\,km\,s$^{-2}$, corresponding to the case in which the solar surface gravity $g_\odot=0.27$\,km\,s$^{-2}$ and the jet's trajectory inclined at 50 degrees. The applied optical flow method is able to calculate the horizontal velocity field of the numerous fine and dynamical structures as seen in this figure.

\begin{figure*}[tb]
\begin{center}
\FigureFile(170mm,76mm){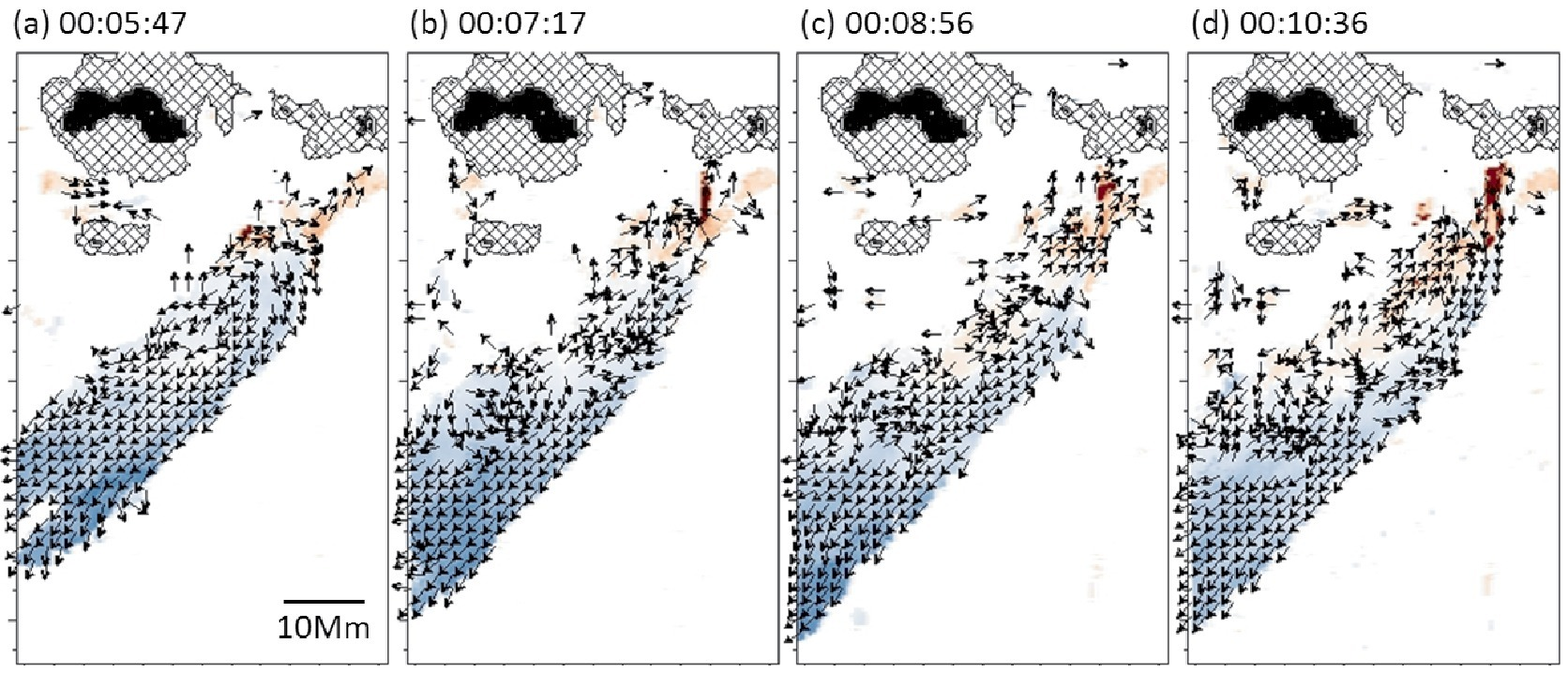}
\caption{The horizontal velocity field measured from the AIA 304\AA\ images by the optical flow method, overlaying the line-of-sight velocity field. The directions of many of arrows in the red-shift region are opposite with that in the blue-shift region. (Color online)}
\label{fig:hv_losv}
\end{center}
\end{figure*}
\begin{figure}[tb]
\begin{center}
\FigureFile(45mm,61mm){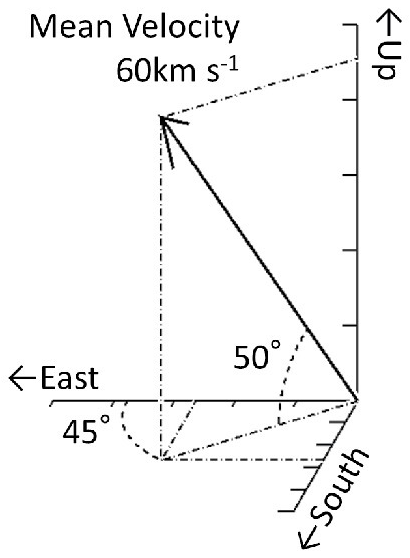}
\caption{The polar coordinated system view of the observed jet at 00:06. The arrow represents the mean velocity vector of the whole of jet in the three-dimensional velocity space. Its elevation is 50 degrees, and azimuth is 45 degrees from east. Each interval on the scale represents 10\,km\,s$^{-1}$.}
\label{fig:v3d}
\end{center}
\end{figure}
\subsubsection{Result}
Figure \ref{fig:hv_losv} and \ref{fig:v3d} are the results of this analysis. Figure \ref{fig:hv_losv} is the time sequence of horizontal velocity field which is represented by the arrows and overlaying on the line-of-sight velocity field. Figure \ref{fig:v3d} shows the polar coordinated system view of the observed jet at 00:06, where the arrow represents the mean velocity vector of the whole jet in the three-dimensional velocity space. The jet was directed to the southeast, and ascended at 50 degrees from the solar surface. \par

\begin{figure*}[tbp]
\begin{center}
\FigureFile(170mm,132mm){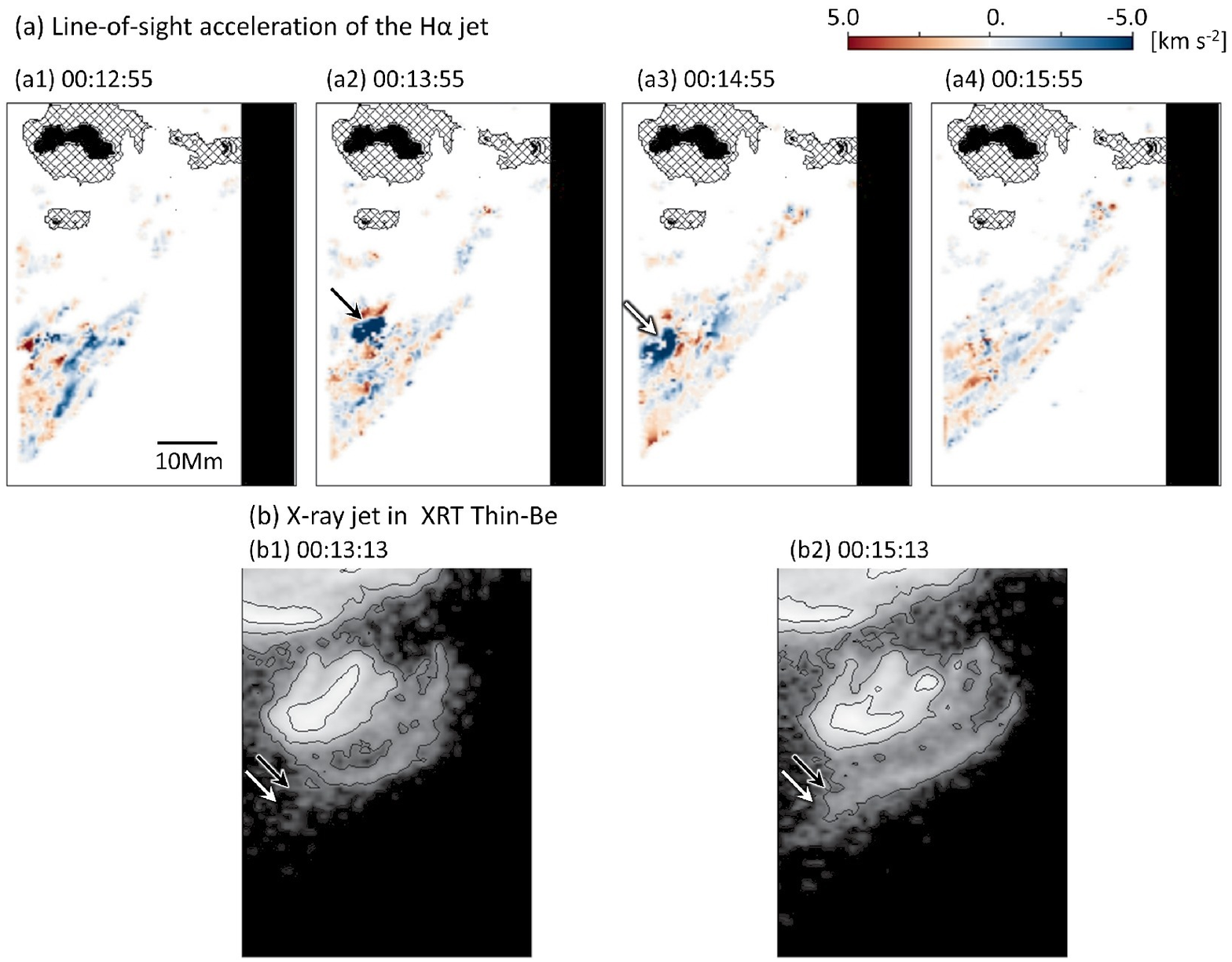}
\caption{Time sequence of the line-of-sight acceleration field of the H$\alpha$ jet (panels (a)) and the XRT images (panels (b)). At 00:13:55 and 00:14:55, the strong upward acceleration of H$\alpha$ jet was observed as indicated with the black and white arrows in panels (a), which are also overlaid on the XRT images in panels (b). The acceleration region lay in the vicinity of the intersection between H$\alpha$ jet and X-ray jet. This result indicates the causal relation between the secondary acceleration of the H$\alpha$ jet and the eruption of the X-ray jet. (Color online)}
\label{fig:acc_xrt}
\end{center}
\end{figure*}

On the basis of the obtained three-dimensional velocity components, we derived the line-of-sight acceleration field of the H$\alpha$ jet. Figure \ref{fig:acc_xrt} is the time sequence of the line-of-sight acceleration field of the H$\alpha$ jet, along with the XRT images in the same period of time. As shown in this figure, although the magnitude of the line-of-sight acceleration are basically a few km\,s$^{-2}$ at most, the stronger upward acceleration region with $\sim$15\,km\,s$^{-2}$ appears at 00:13:55 and 00:14:55, which correspond to the period when the line-of-sight velocity of the H$\alpha$ jet re-increased (figure \ref{fig:vlos_time}). These localized acceleration regions (hereafter called ``secondary acceleration regions") are indicated with the black and white arrows in panels (a), and overlaid on the XRT images. It is remarkable that these regions are in the vicinity of the intersection between H$\alpha$ jet and X-ray jet.

Therefore, we conclude that the re-increase of line-of-sight velocity in figure \ref{fig:vlos_time} is accounted for by the additional acceleration of a part of ejecta in H$\alpha$ jet after it had been ejected from the lower atmosphere.

\subsection{Investigation of the Secondary Acceleration Region}
\label{sec:Investigation of the Secondary Acceleration Region}

\begin{figure*}[tbp]
\begin{center}
\FigureFile(170mm,160mm){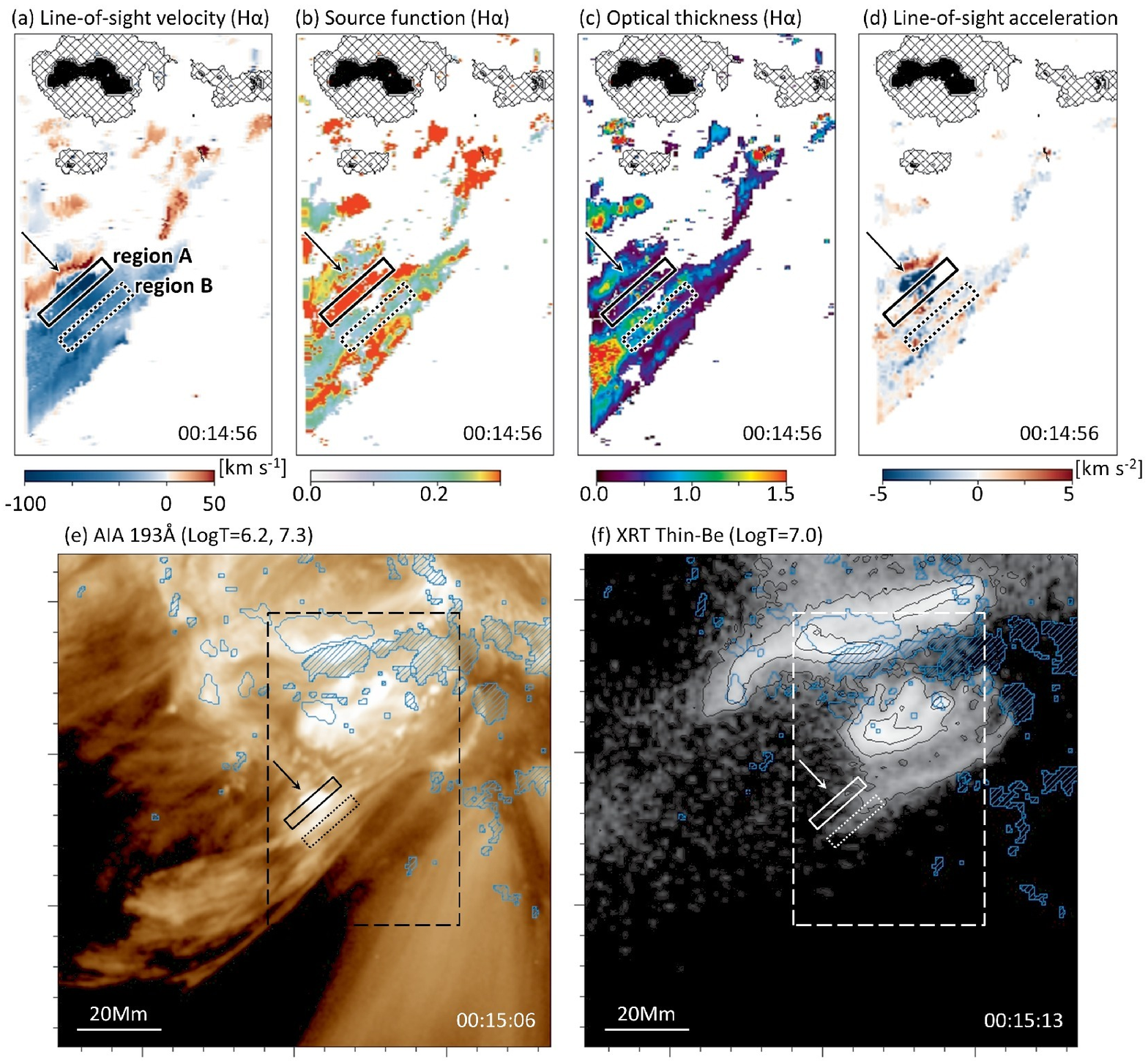}
\caption{The multi-wavelength images and physical quantities distributions
around the secondary acceleration region at 00:15 observed in DST H$\alpha$ [panels (a)--(d)], AIA 193\AA\ [panel (e)], and XRT image [panel (f)]. The black contours in panel (f) show the X-ray brightness distribution, and blue ones in panels (e, f) represent the regions where the absolute magnetic field strength is larger than 200G; also, the negative polarities are shaded. The field of view of DST images is represented by the dash-line frame in panels (e) and (f), and that of panels (e) and (f) is the same as figures \ref{fig:ha+aia304time_sequence}--\ref{fig:xrt_time_sequence}. The characteristic temperatures of these panels range from $10^4$ K to $10^7$ K. The secondary acceleration region at 00:15 is indicated with the arrows. Around that region, the brightening blob is seen in 193\AA\ image, and the source function and optical thickness of H$\alpha$ are respectively larger and smaller than the surrounding. The temporal variations of physical quantities in the region labeled region A and B are presented in figure \ref{fig:acc_temp}. (Color online)}
\label{fig:sar}
\end{center}
\end{figure*}

Here, we provide some results of the additional analysis with the multi-wavelength data on the secondary acceleration regions which are specified in the previous subsection. Figure \ref{fig:sar} shows the multi-wavelength images and physical quantities distributions around the secondary acceleration region at 00:15 observed in DST H$\alpha$, AIA 193\AA, and XRT image. Their characteristic temperatures range from $10^4$ K to $10^7$ K. The secondary acceleration region is indicated with the arrows. The most remarkable feature in this figure is the brightening blob in AIA 193\AA\ image which lay around the secondary acceleration region. The figure also shows that the source function and optical thickness of H$\alpha$ in the region are respectively larger and smaller than the surrounding, as if the region laid on the ridge of the source function field, or at the trough of the optical thickness. These features suggest the possibility that the secondary acceleration region was full of the plasma with high temperature enough to be observed in EUV.

To realize when and how the hot plasma had increased the presence in the H$\alpha$ jet, we compare the temporal variations of the physical quantities in the two regions enclosed with the solid- and dash-line rectangles in figure \ref{fig:sar}. These regions are labeled region A and B; the former corresponds to the secondary acceleration region. 

\begin{figure*}[tbp]
\begin{center}
\FigureFile(136mm,180mm){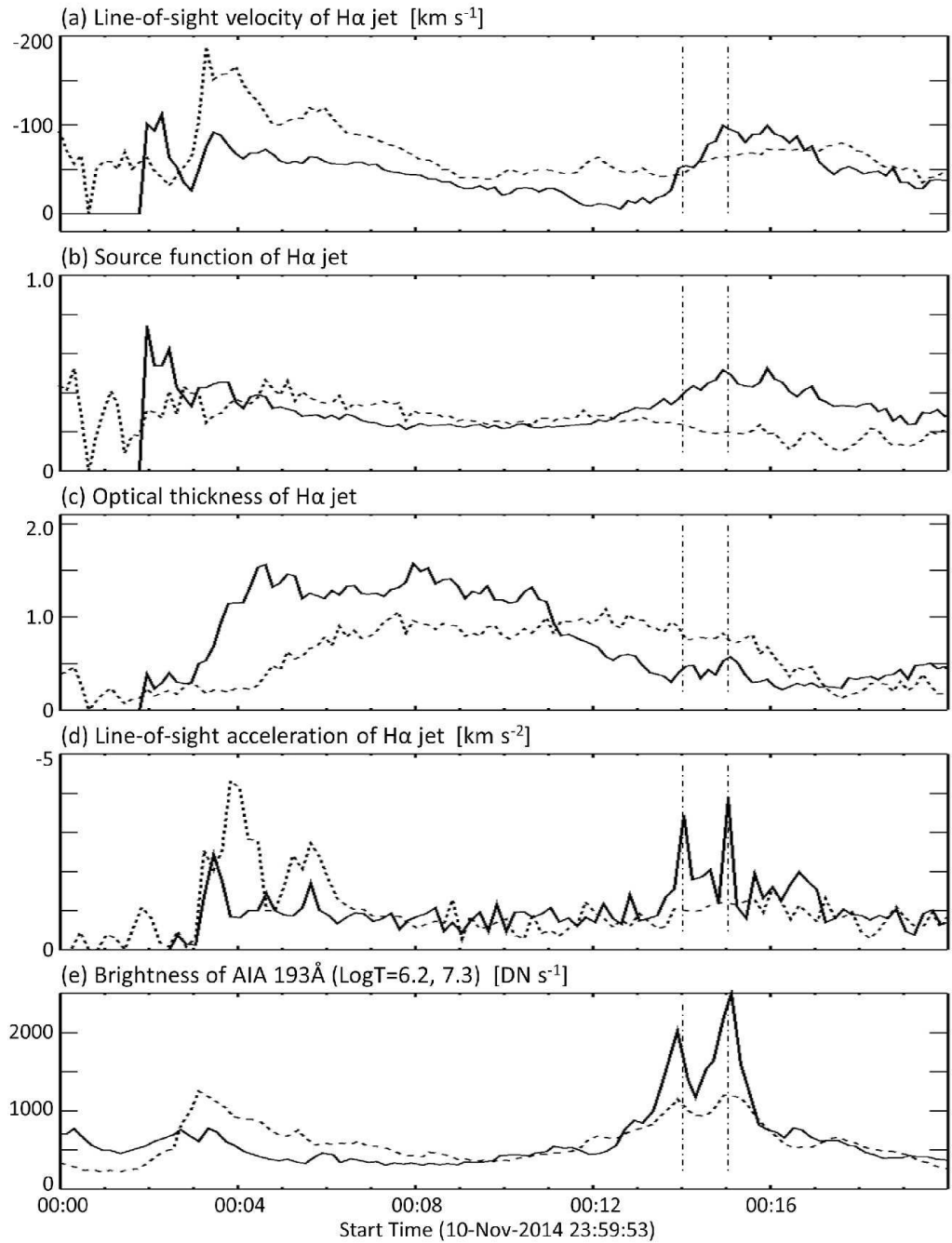}
\caption{Temporal variations of (a) line-of-sight velocity, (b) source function, (c) optical thickness, (d) line-of-sight acceleration of H$\alpha$ jet, and (e) brightness of AIA 193\AA, which are averaged over the region A and B, respectively (see figure \ref{fig:sar}). The region A represents the secondary acceleration region, and the dash-dot lines correspond to the times when it occurred (figure \ref{fig:acc_xrt}). The panels (d) and (e) clearly shows the secondary acceleration of H$\alpha$ jet synchronized with the impulsive enhancements of the AIA 193\AA\ in the same region. Moreover, the source function [panel (b)] and optical thickness [panel (c)] in the region A became significantly higher and lower than those in the region B after 00:10, preceding the secondary acceleration.}
\label{fig:acc_temp}
\end{center}
\end{figure*}

In figure \ref{fig:acc_temp}, the solid and dash lines represent the temporal variations of (a) line-of-sight velocity, (b) source function, (c) optical thickness, (d) line-of-sight acceleration of H$\alpha$ jet, and (e) brightness of AIA 193\AA, which are averaged over the region A and B, respectively. Note that the components with the downward acceleration of H$\alpha$ jet is avoided in the calculation of the average in panel (d), so as to highlight the strong upward acceleration seen in figure \ref{fig:acc_xrt}. 

The mentioned strong upward acceleration of H$\alpha$ jet occurred at the times indicated by the dash-dot lines in this figure, and the panels (d) and (e) clearly show that they synchronized with the impulsive enhancements of the AIA 193\AA\ intensity in the same region. Moreover, the source function [panel (b)] and optical thickness [panel (c)] in the region A became significantly higher and lower than those in the region B after 00:10, preceding the secondary acceleration. These trends represent the formation of the ridge in the source function field and trough in the optical thickness field of H$\alpha$.

\section{Discussion}
\label{sec:discussion}
\begin{figure*}[tb]
\begin{center}
\FigureFile(120mm,60mm){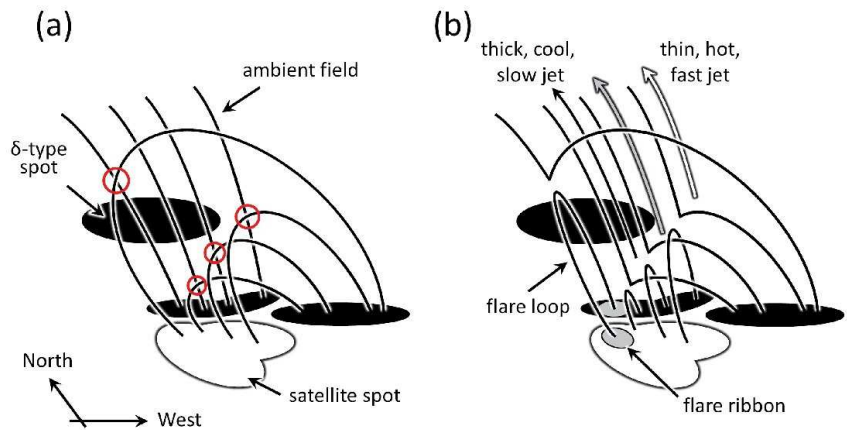}
\caption{The schematic image for the mechanism to form the gradient of the distribution of the physical quantities of H$\alpha$ jet. As the footpoint of jet migrated from east to west (figure \ref{fig:jet_timeslice_iris}), the height of reconnection site, indicated with the red circles, became higher, and then the temperature, density, and velocity of the jet got higher, thinner, and faster respectively. Consequently, the source function, optical thickness, and line-of-sight velocity of the H$\alpha$ jet increase from east to west, which are observed as the observed internal structure.}
\label{fig:jet_grad}
\end{center}
\end{figure*}

\begin{figure*}[tb]
\begin{center}
\FigureFile(170mm,119mm){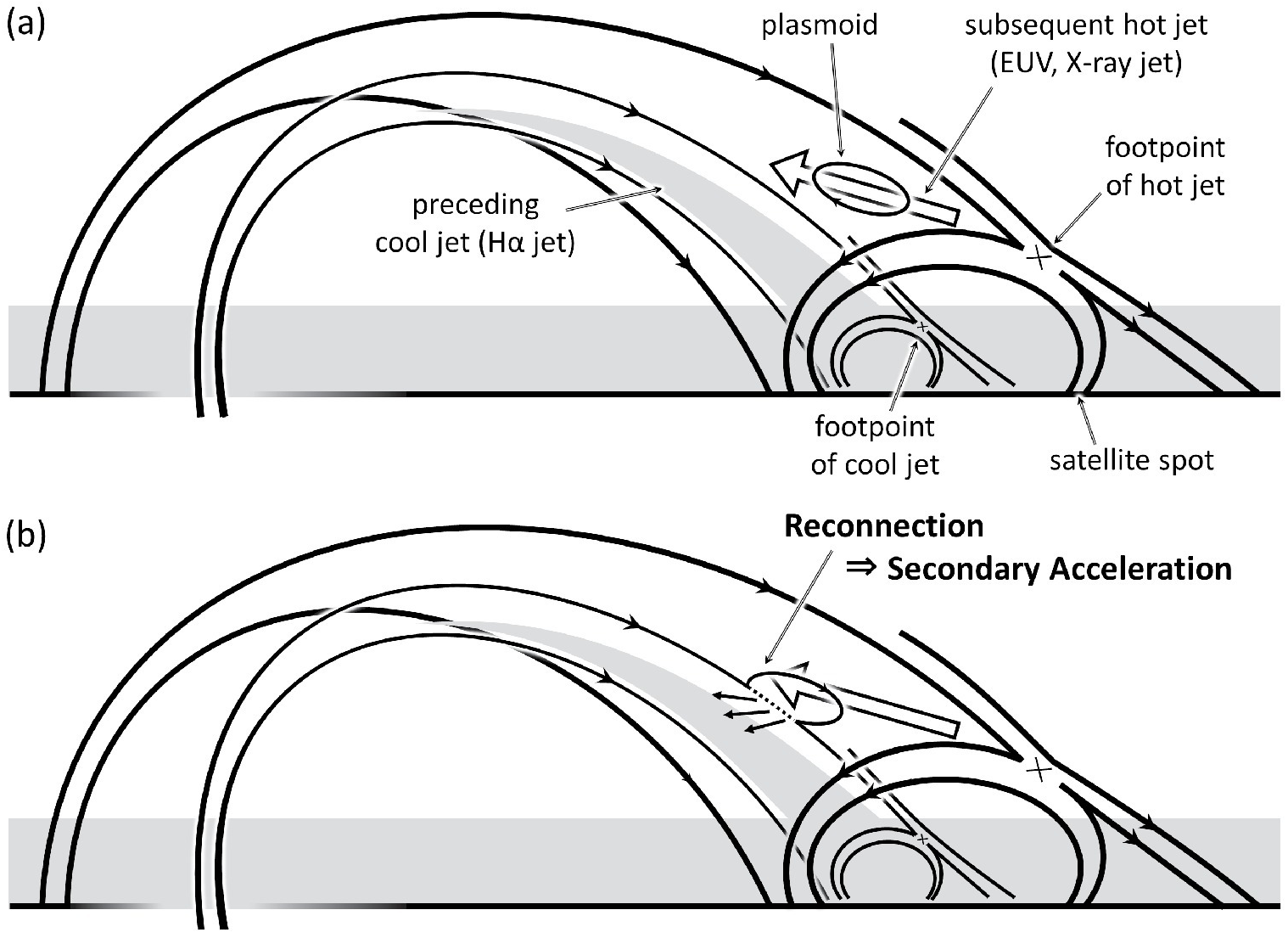}
\caption{The schematic image for the mechanism of the secondary acceleration. We speculate the reconnection occurred between the magnetic flux hosting the subsequent hot jet with that of the preceding cool jet. The reconnection caused the heating of the cool plasma, and the effective sound speed in the jet became larger, which means the acceleration of the jet was facilitated. In particular, the impact of the X-ray jet to the H$\alpha$ jet would lead to the stronger acceleration. That is how the secondary acceleration occurred.}
\label{fig:secondary_acc}
\end{center}
\end{figure*}

In this section, we discuss the following two analyses results, especially in terms of the three-dimensionality of the observed jet. First is the origin of the gradient of physical quantities in H$\alpha$ jet (seen in figure \ref{fig:vlos_grad}), and second is the mechanism of the secondary acceleration in the vicinity of the intersection between the H$\alpha$ jet and X-ray jet (seen in figure \ref{fig:vlos_time} and figure \ref{fig:acc_xrt}). \par

\subsection{The gradient of physical quantities in H$\alpha$ jet}
From the detailed discussion in Paper I, we conjecture the coronal field configuration around the jet ejection site as depicted in figure \ref{fig:jet_grad}, and this figure is also a schematic image for the formation of the gradient of physical quantities in H$\alpha$ jet. According to the unified model (section 1), the relative height of the energy release site (reconnection site) to the transition layer determines the properties of the associated explosive phenomena, including the cool jet ($\sim 10^4$K), hot jet ($\sim 10^{6-7}$K), and flare ($\sim 3\times10^7$K). Thus, it is reasonable that the individual fine threads of the jet are characterized by the different physical quantities, corresponding to the heights of their energy release sites.\par

In the case of the observed phenomena, figure \ref{fig:jet_timeslice_iris} shows the footpoint of cool jet in IRIS images gradually moved from east to west, where the footpoint of hot jet was. This suggests the energy release site of the cool jet migrated from east to west and relatively lower atmosphere to higher atmosphere as indicated with the red circle in figure \ref{fig:jet_grad}. As this migration of jet footpoint, the temperature and density of jet became higher and smaller respectively, which could lead to the observed gradient in the east-west direction of the source function and optical thickness of H$\alpha$ jet. The initial launch velocity of ejecta, similarly, had been increasing with time, and it is natural that the falling flow, seen as the red shift in the east side of jet, developed alongside with the ascending flow corresponding to the blue-shift component. Note that we do not interpret the separation of the line-of-sight velocity field of H$\alpha$ jet as the result of some spinning motion. Actually, in figure \ref{fig:hv_losv}, many of arrows representing the horizontal velocity vectors in the red-shift region are opposite with those in the blue-shift region, which supports the above statement.  \citet{1973PASJ...25..447T} also reported the both upward blue and downward redshift components in the H$\alpha$ jet at the same time and interpreted it as the result of the difference in their initial ejection velocities.

As for the gradient of the velocity dispersion field, we interpret it is simply resulted from the braiding of jet threads. Since the observed jet traversed along the closed magnetic loop from west to east while its footpoint moved from east to west, it is likely that the jet threads were braided. Consequently, the subsequent jet threads approached the preceding ones from the west side, and overlapped them in the line-of-sight direction. In the profile of spectrum, that was observed as the mixing of several components with the different line-of-sight velocities, which means the large velocity dispersion in the line-of-sight direction.

\subsection{The secondary acceleration of the H$\alpha$ jet}
The second important result of our analyses is to specify that a part of ejecta in the H$\alpha$ jet experienced the additional acceleration after it had been ejected from the lower atmosphere. This phenomenon are characterized as follows. First, this secondary acceleration occurred in the vicinity of the intersection between the H$\alpha$ jet and the subsequent X-ray jet. Second, it was associated with the appearance of the brightening blob in the AIA 193\AA\ image, the ridge-like feature in the source function field of H$\alpha$ and the trough-like feature in its optical thickness field. From these observed features, we speculate the three-dimensional configuration of the jet as shown in figure \ref{fig:secondary_acc}. This schematic drawings depict that the emitted plasmoid in the subsequent hot jet collided against the preceding cool jet. At this instance, the magnetic field enclosing the plasmoid reconnected with the magnetic flux hosting the cool jet, and the released magnetic energy was converted to the kinetic or thermal energy of the ejecta in situ. It was observed as the appearance of the brightening blob in the AIA 193\AA\ image, and as the onset of re-increase in the line-of-sight velocity of H$\alpha$ jet (figure \ref{fig:vlos_time}). As the result of so effective heating that the neutral hydrogens were ionized, there was formed the flow along the jet in which the source function and optical thickness of H$\alpha$ was large and small, respectively. It was observed as the ridge-like feature in the source function field of H$\alpha$ and trough-like feature in the optical thickness field. Moreover, as the filling factor of the cool ejecta became smaller in the jet, the additional acceleration of cool jet was facilitated. In particular, the impact of the X-ray jet to the H$\alpha$ jet would cause the stronger acceleration, which was confirmed in figure \ref{fig:acc_xrt}. That is how the secondary acceleration occurred. \par

Note that \citet{1999Ap&SS.264..129S} proposed the similar magnetic configuration for the untwisted jet eruption which is induced by the reconnection between the plasmoid with the ambient loop, and \citet{2007PASJ...59S.745S} observed the features in the X-ray jet. Our and their studies suggest that the properties of jet are possible to be transfigured by these reconnections away from the jet footpoint, which causes the secondary acceleration in our case.\par

Unfortunately, there are not enough observational evidence for the above speculation, especially for the heating mechanism, but the energetics is consistent at least as mentioned below. The magnetic energy released by the reconnection is estimated as:
\[
E_{\mbox{\scriptsize M}}=f{B^2\over8\pi}L_{\mbox{\scriptsize SAR}}^3=10^{26}\mbox{ erg}\times\left({f\over0.5}\right)\left({B\over 10\mbox{ G}}\right)^2\left({L_{\mbox{\scriptsize SAR}}\over5000\mbox{ km}}\right)^3
\nonumber
\]
where $f$ is the fraction of energy release and $L_{\mbox{\scriptsize SAR}}$ is the spatial scale length of the secondary acceleration region. On the other hand, the kinetic energy and thermal energy, contributing to the acceleration and the ionization of the H$\alpha$ jet are estimated as:
\begin{eqnarray}
E_{\mbox{\scriptsize K}}\! &=&\! {1\over2}\rho(\Delta v)^2L_{\mbox{\scriptsize SAR}}^3=4\times10^{25}\mbox{ erg} \nonumber \\
\! \! &&\! \! \times\left({n\over10^{10}\mbox{ cm}^3}\right)\left({\Delta v\over 60\mbox{ km s}^{-1}}\right)^2\left({L_{\mbox{\scriptsize SAR}}\over5000\mbox{ km}}\right)^3
\nonumber \\
E_{\mbox{\scriptsize T}}\! &=&\! {3\over2}\rho R\Delta TL_{\mbox{\scriptsize SAR}}^3
=3\times10^{25}\mbox{ erg}\nonumber \\
\! \! &&\! \! \times\left({n\over10^{10}\mbox{ cm}^3}\right)\left({T_{\mbox{\scriptsize ionized}}\over 10^5\mbox{ K}}\right)\left({L_{\mbox{\scriptsize SAR}}\over5000\mbox{ km}}\right)^3
\nonumber
\end{eqnarray}
where $R$ is the gas constant,  $\Delta v$ is the increase in velocity due to the secondary acceleration, which is estimated from Fig\,\ref{fig:vlos_time}, and $T_{\mbox{\scriptsize ionized}}$ is the temperature at which the H$\alpha$ jet is fully ionized. These suggest the possibility that the released energy by the reconnection was converted to both kinetic energy and thermal energy of part of ejecta in H$\alpha$ jet in equipartition. Note that those quantities are small fractions of the total energy involved in the H$\alpha$ jet ejection, which is estimated as the potential energy stored in jet when it reached the highest altitude: $E_{\mbox{\scriptsize total}}={1\over2}\rho g_\odot AH^2\sim4\times10^{28}\mbox{erg}$. Here, $\rho\sim10^{-14}$g cm$^{-3}$ is the mass density of jet, $g_\odot=2.74\times10^4$cm s$^{-2}$ is the solar surface gravity. $A$ and $H$ are the cross section ($\sim\pi\times(10\mbox{Mm})^2$) and the highest altitude ($\sim100$Mm) of jet, respectively. This total energy is typical value for the H$\alpha$ jet ejection (\cite{1996SoPh..168...91K}; \cite{2004ApJ...610.1136L}).
\par

\section{Summary}
We analyzed the jet phenomenon which is associated with the C5.4 class flare on 2014 November 11. The data of the jet is provided by SDO, Hinode/XRT, IRIS and DST of Hida Observatory. The spectral data observed by DST enabled us to obtain the spatial distribution of the physical quantities of the H$\alpha$ jet, including the source function, optical thickness, line-of-sight velocity, and velocity dispersion, by spectral analysis based on the cloud model. Additional analyses on the horizontal velocity field of the jet by using the optical flow method made it possible to discuss the line-of-sight acceleration field of the jet. These analyses revealed the following two.\par

(1) In the physical quantity distributions of the H$\alpha$ jet, there is a significant gradient in the direction crossing the jet. The optical thickness in the right side as viewed toward the travel direction of the jet was larger than left side, while the source function and velocity dispersion in the right side was smaller than left side. As for the line-of-sight velocity field, that in the right side was smaller than the left side, and the red shift had gradually developed in the right side. These phenomena are basically interpreted as the result of the migration of the energy release site from relatively lower atmosphere to relatively upper atmosphere. \par
(2) A part of ejecta in the H$\alpha$ jet experienced the secondary acceleration after it had been ejected from the lower atmosphere. We specified the secondary acceleration region in the vicinity of the intersection between the trajectories of H$\alpha$ jet and X-ray jet, and inferred a fundamental cause of it is the magnetic reconnection between the preceding H$\alpha$ jet with plasmoid in the subsequent X-ray jet. We also observed the heating of the cool plasma around the secondary acceleration region, and discussed the magnetic energy would be converted to the kinetic and thermal energy in equipartition. 
\\
\par
\begin{ack}
We want to thank Dr. K. Otsuji for his developing flat-field reduction method for the spectral images taken by the spectrograph of DST. Hinode is a Japanese mission developed and launched by ISAS/JAXA, with NAOJ as domestic partner and NASA and UKSA as international partners. It is operated by these agencies in co-operation with ESA and NSC (Norway). HMI and AIA are instruments on board SDO, a mission for NASA's Living With a Star program. IRIS is a NASA small explorer mission developed and operated by LMSAL with mission operations executed at NASA Ames Research center and major contributions to downlink communications funded by ESA and the Norwegian Space Centre. This work was supported by JSPS KAKENHI Grant Numbers JP16H03955, JP15H05814, and JP15K17772. AA is supported by a Shiseido Female Researcher Science Grant. 
\end{ack}

\appendix
\section{modification of contrast function}
\label{sec:modification of contrast function}
\begin{figure*}[tb]
\begin{center}
\FigureFile(170mm,107mm){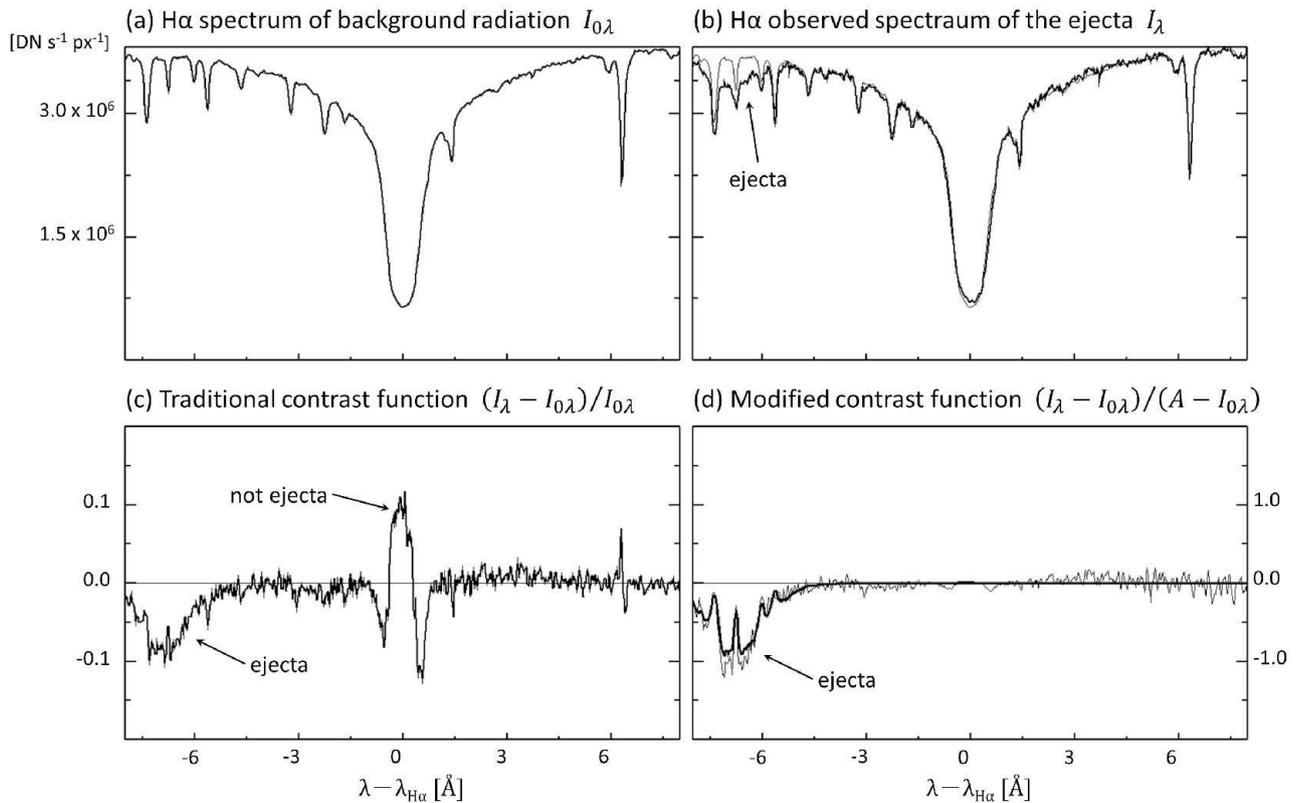}
\caption{(a) The spectrum ($I_\lambda$) in which the absorption signal shown in the figure \ref{fig:dst_his_hs} appears. (b) The presumed spectrum ($I_{0\lambda}$) of background radiation. (c) The traditional contrast function defined by $I_\lambda$ and $I_{0\lambda}$ as $(I_\lambda-I_{0\lambda})/I_0$. (d) The modified contrast function defined by $I_\lambda$, $I_{0\lambda}$ and a constant value $A$ as $(I_\lambda-I_{0\lambda})/(A-I_{0\lambda})$ (thin line), and the fitted function (thick line). Note that $A$ is determined empirically to $1.05\times\mbox{continuum level of }I_{0\lambda}$.}
\label{fig:25-40-172}
\end{center}
\end{figure*}
We explain the details of our spectral analysis in the following appendix 1 and 2, especially here the modification of the contrast function in the cloud model. The panels (a), (b), (c) of figure \ref{fig:25-40-172} show the observed H$\alpha$ spectrum, presumed background spectrum, and the traditional contrast function $(I_\lambda-I_{0\lambda})/I_{0\lambda}$. The background profile is defined by averaging the spectra for the periods when the ejecta does not affect the radiation from the lower atmosphere. The observed spectrum in figure \ref{fig:25-40-172} corresponds to a cross section of the spectral image in the panel (b) of figure \ref{fig:dst_his_hs}, and the signal of ejecta, pointed at by the arrow in that figure, appears as an absorption in the blue continuum of the observed spectrum.\par

The necessity of the modified contrast function arises from the problem that the traditional contrast function in figure \ref{fig:25-40-172} comprises not only the signal in the wing, which represents the absorption by the ejecta, but also that in the line core, which is probably produced by some fine dynamics of the lower atmosphere. Because the latter is not our present interest, we intended to ignore the signal in the line core, and thus, we modified the contrast function with an arbitrary constant value $A$ as follows:
\begin{equation}
{I_\lambda-I_{0\lambda}\over A-I_{0\lambda}}={S-I_{0\lambda}\over A-I_{0\lambda}}(1-e^{-\tau_\lambda}) \label{eq:modified}
\end{equation}
where $\tau_\lambda$ is described as Eq.\,(\ref{eq:tau}). We empirically took $1.05\times\mbox{continuum level of }I_{0\lambda}$ for $A$. The panel (d) of figure \ref{fig:25-40-172} shows the modified contrast function of the spectra (thin line) and the fitted function (thick line) with the obtained parameters; $S$ of 0.3\% of continuum level, $\tau_0$ of 0.10, $v_{\mbox{\scriptsize los}}$ of $-$314\,km\,s$^{-1}$, $W$ of 30\,km\,s$^{-1}$.\par

This modification makes the signal in the line core much smaller, in spite of the mathematical equivalence between Eq.\,(\ref{eq:modified}) and Eq.\,(\ref{eq:cloud_model}), and enables our spectral analysis, especially our algorithm for automatically fitting, to focus on more wavelength-shifted signals.\par
Note that there is a drawback to the modified contrast function, which is that the non-uniformity of the sensitivity of CCD is not removed. Due to it, the observed value $I_\lambda$ usually deviates from the true value $\bar{I}_\lambda$ by the factor $f_\lambda$ (i.e. $I_\lambda=\bar{I}_\lambda f_\lambda$). While the factor $f_\lambda$ is canceled out in the traditional contrast function,
 \[\mbox{i.e.}\ \ \ \ \ (I_\lambda-I_{0\lambda})/I_{0\lambda}=(\bar{I}_\lambda-\bar{I}_{0\lambda})/\bar{I}_{0\lambda}\]
that remains in the modified contrast function.

\section{overlap of ejectas in the line-of-sight direction}
\label{sec:overlap of ejectas in the sight line direction}

\begin{figure*}[tb]
\begin{center}
\FigureFile(170mm,56mm){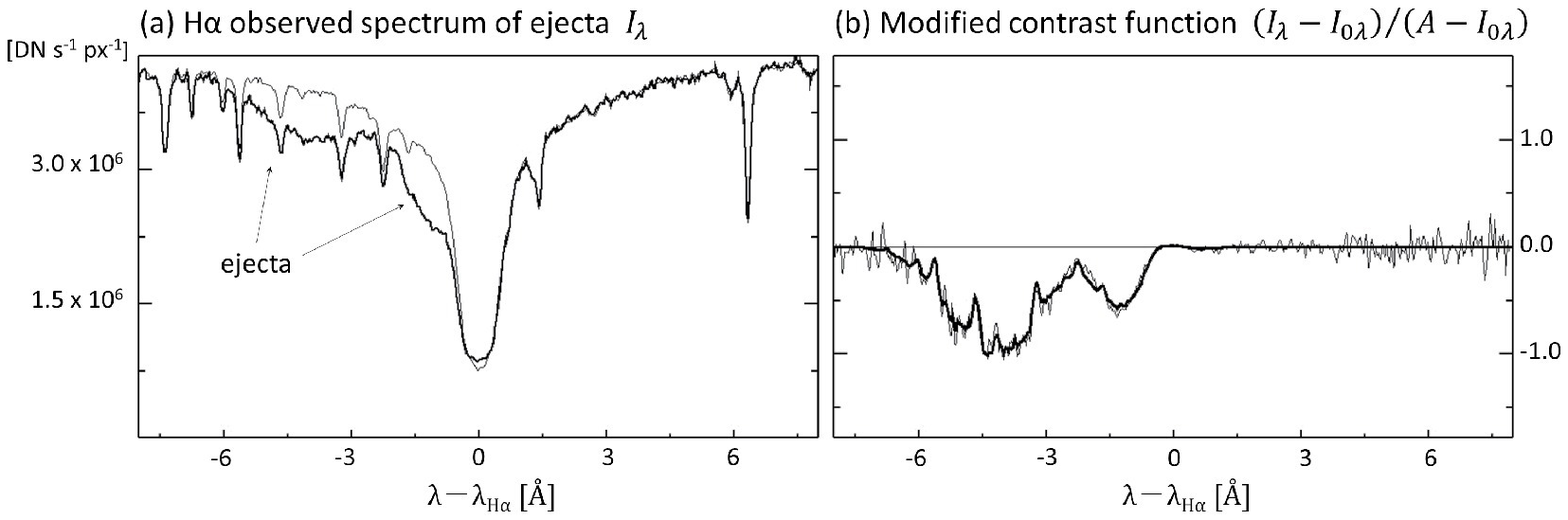}
\caption{(a) The spectrum in which a separation of the absorption profile appears (thick line), and the spectrum of the background radiation (thin line). (b) the modified contrast function defined by the spectra shown in the left panel (thin line), and the fitting result with two cloud model (thick line)}
\label{fig:2c_50-20-160}
\FigureFile(170mm,80mm){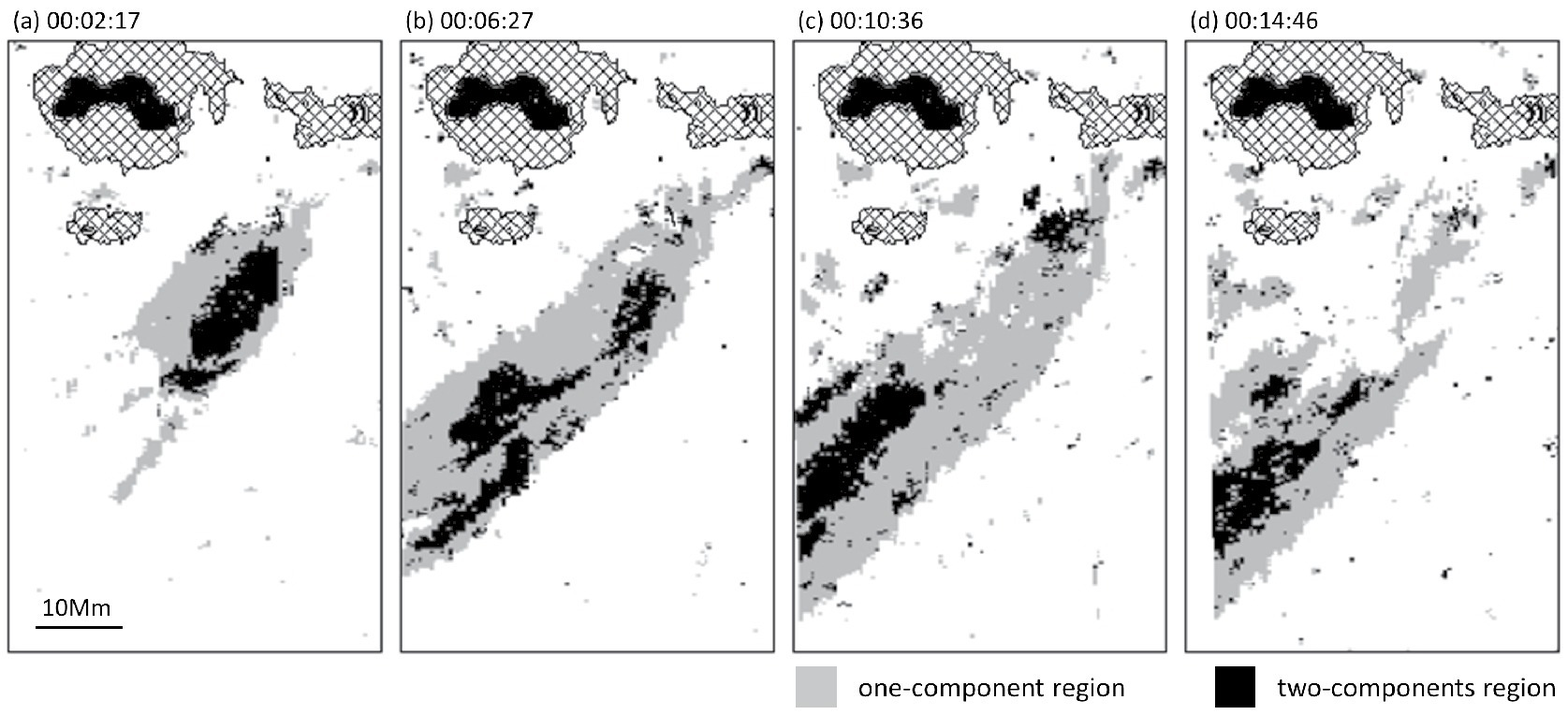}
\caption{The spatial distribution of the number of components in the profile of contrast function (see figure \ref{fig:2c_50-20-160}), and its time sequence in an interval of 250 seconds. The two-components regions (colored with black) appear to form into a lump surrounded by the one-component regions (colored with gray).}
\label{fig:2com_map}
\end{center}
\end{figure*}

Secondly, we explain the way to handle such spectra as shown in figure \ref{fig:2c_50-20-160}, in which the absorption profile is separated into two components. This separation indicates the ejecta has a two-layer structure which have different velocities and overlap each other in the line-of-sight direction. For these spectra, therefore, we used the two cloud model \citep{1992A&A...259..649G}, in which the modified contrast function is generalized as below:
\begin{eqnarray}
{I_\lambda-I_{0\lambda}\over A-I_{0\lambda}}\! \! &=&\! \!
{I_{0\lambda}\over A-I_{0\lambda}}\left[
e^{-\tau^u_\lambda-\tau^l_\lambda}-1\right.\nonumber \\
\! \! &+&\! \! \left.
{S^u\over I_{0\lambda}}(1-e^{-\tau_\lambda^u})  \right.\nonumber \\
\! \! &+&\! \! \left.
{S^l\over I_{0\lambda}}(1-e^{-\tau_\lambda^l})e^{-\tau_\lambda^u}\right]
\end{eqnarray}

\begin{equation}
\tau_\lambda ^j=\tau_{0}^j\exp\left[-{1\over2}\left({\lambda/\lambda_0-(1+v_{\mbox{\scriptsize los}}^j/c)\over W^j/c}\right)^2\right] \label{eq:2ctau}
\end{equation}
($j=u$ or $l$), where the subscript $u$, $l$ represent the upper and lower cloud, each of which is characterized by the four parameters; $S^j$, $\tau_0^j$, $v_{\mbox{\scriptsize los}}^j$, $W^j$, as well as in the case of one cloud model. In the panel (b) of figure \ref{fig:2c_50-20-160}, the result of the two cloud model fitting is drawn with the thick line. Figure \ref{fig:2com_map} shows the areas where the two components are distinguished in the spectra (hereafter called ``two-components region"), and their temporal evolution. From this figure, it is clear that the two-layer structure is widely distributed in the observed jet, which let us imagine the three-dimensional structure of the jet. It depends on the degree of the separation between the components, which of one or two cloud model is proper. That means the two-layer structure does not always lead to such a separation in the profile, although the separation in the profile of contrast function is always the evidence of a two-layer structure. Finally we should note that the value of line-of-sight velocity in the two-components region displayed in figure \ref{fig:vlos_ev} and figure \ref{fig:vlos_time} are synthesized from the values of upper and lower cloud as $v_{\mbox{\scriptsize los}}:=(v_{\mbox{\scriptsize los}}^u\tau^u+v_{\mbox{\scriptsize los}}^l\tau^l)/(\tau^u+\tau^l)$. Similarly, the other physical quantities in the two-components region displayed in the following are synthesized as $S:=(S^u\tau^u+S^l\tau^l)/(\tau^u+\tau^l)$, $\tau:=\tau^u+\tau^l$, and $W:=W^u+W^l$, respectively. \par

\end{document}